\documentclass[a4paper,fleqn,usenatbib]{mnras}

\usepackage{newtxtext,newtxmath}

\usepackage[T1]{fontenc}
\usepackage{ae,aecompl}

\usepackage{graphicx}	
\usepackage{amsmath}	
\usepackage{amssymb}	
\usepackage{multirow}

\newcommand\atd{ATLAS$^{\rm 3D}$}
\newcommand\mh{M_{\rm halo}}
\newcommand\kms{km s$^{-1}$}
\def\gsim{\;\rlap{\lower 2.5pt \hbox{$\sim$}}\raise 1.5pt\hbox{$>$}\;} 
\def\lsim{\;\rlap{\lower 2.5pt \hbox{$\sim$}}\raise 1.5pt\hbox{$<$}\;}

\newcommand{\aref}[1]{\mbox{\hyperref[#1]{Appendix~\ref{#1}}}}
\newcommand{\mautoref}[1]{\mbox{\autoref{#1}}}

\title[MASSIVE VII. - Environment and Angular Momentum]{The MASSIVE Survey - VII. The Relationship of Angular Momentum, Stellar Mass and Environment of Early-Type Galaxies}

\author[Veale et al.]{
  Melanie Veale,$^{1,2}$\thanks{E-mail: melanie.veale@berkeley.edu (MV), cpma@berkeley.edu (C-PM)}
  Chung-Pei Ma,$^{1,2}$\footnotemark[1]
  Jenny E. Greene,$^3$
  Jens Thomas,$^4$
  \newauthor
  John P. Blakeslee,$^5$
  Nicholas McConnell,$^5$
  Jonelle L. Walsh,$^6$
  Jennifer Ito$^1$
\\
$^{1}$Department of Astronomy, University of California, Berkeley, CA 94720, USA \\
$^{2}$Department of Physics, University of California, Berkeley, CA 94720, USA \\
$^3$Department of Astrophysical Sciences, Princeton University, Princeton, NJ 08544, USA \\
$^4$Max Plank-Institute for Extraterrestrial Physics, Giessenbachstr. 1, D-85741 Garching, Germany \\
$^5$Dominion Astrophysical Observatory, NRC Herzberg Astronomy and Astrophysics, Victoria BC V9E2E7, Canada \\
$^6$George P. and Cynthia Woods Mitchell Institute for Fundamental Physics and Astronomy, and Department of Physics and Astronomy, \\Texas A\&M University, College Station, TX 77843, USA \\
}

\date{Accepted XXX. Received YYY; in original form ZZZ}

\pubyear{2017}

\begin{document}
\label{firstpage}
\pagerange{\pageref{firstpage}--\pageref{lastpage}}
\maketitle

\begin{abstract}

We analyse the environmental properties of 370 local early-type galaxies (ETGs) in the MASSIVE and \atd\ surveys, two complementary volume-limited integral-field spectroscopic (IFS) galaxy surveys spanning absolute $K$-band magnitude $-21.5 \ga M_K \ga -26.6$, or stellar mass $8\times 10^{9} \la M_* \la 2\times 10^{12} M_\odot$.
We find these galaxies to reside in a diverse range of environments measured by four methods: group membership (whether a galaxy is a brightest group/cluster galaxy, satellite, or isolated), halo mass, large-scale mass density (measured over a few Mpc), and local mass density (measured within the $N$th neighbour).
The spatially resolved IFS stellar kinematics provide robust measurements of the spin parameter $\lambda_e$
and enable us to examine the relationship among $\lambda_e$, $M_*$, and galaxy environment.
We find a strong correlation between $\lambda_e$ and $M_*$, where the average $\lambda_e$ decreases from $\sim 0.4$ to below 0.1 with increasing mass, and the fraction of slow rotators $f_{\rm slow}$ increases from $\sim 10$\% to 90\%.
We show for the first time that at fixed $M_*$, there are almost no trends between galaxy spin and environment; the apparent kinematic morphology-density relation for ETGs is therefore primarily driven by $M_*$ and is accounted for by the joint correlations between $M_*$ and spin, and between $M_*$ and environment.
A possible exception is that the increased $f_{\rm slow}$ at high local density is slightly more than expected based only on these joint correlations.
Our results suggest that the physical processes responsible for building up the present-day stellar masses of massive galaxies
are also very efficient at reducing their spin, in any environment.
\end{abstract}




\section{Introduction}

\label{sec:introduction}

As a group, elliptical galaxies obey the fundamental plane and have predominantly old stellar populations
(e.g. \citealt{dressleretal1987}; \citealt{djorgovskidavis1987,baldryetal2004,thomasetal2005}).
The properties of elliptical galaxies vary considerably with mass, however, and can be grouped into two general families.
Lower-mass elliptical galaxies have flattened isophotes, power-law central light profiles, and some net rotation that aligns with their shortest axis.
More massive elliptical galaxies, in contrast, show boxy isophotes, cored light profiles, and small net rotation
(e.g., \citealt{illingworth1977}; \citealt{daviesetal1987}; \citealt{benderetal1989}; \citealt{kormendybender1996}; \citealt{kormendyetal2009}).
The standard interpretation of these differences is that massive elliptical galaxies experience a large number of relatively gas-free mergers that effectively erase any record of their spin, while at lower mass gas accretion and gas-rich mergers tend to preserve a net spin to the galaxies \citep[e.g.][]{hoffmanetal2010,boisetal2011,moodyetal2014,khochfaretal2011,martizzietal2014,naabetal2014,choiyi2017,penoyreetal2017}.
Stellar mass strongly determines whether a galaxy is a fast or slow rotator \citep[e.g.][]{emsellemetal2011,cappellari2013,vealeetal2017}.
However, given that the merger and accretion history of a galaxy is partially determined by its surrounding large-scale environment, it is also important to assess the impact of galaxy environment on galaxy rotation.

Motivated by the classic morphology-density relation of \citet{dressler1980}, a number of papers have investigated an analagous {\it kinematic} morphology-density relation using integral field spectrograph (IFS) data, comparing how late-type galaxies (LTGs), fast rotating early-type galaxies (ETGs), and slow rotating ETGs populate different density environments.
\citet{cappellarietal2011b} find that substituting kinematic morphology (i.e. fast or slow rotator status) for Hubble type (lenticular versus elliptical) yields a cleaner relationship than the traditional morphology-density relation.
While a significant fraction of elliptical galaxies populate low-density environments, they find that nearly all of them are fast rotators more similar to inclined lenticular galaxies than to genuine spheroidal ellipticals.
They also find that the fraction of slow rotators within the ETG population (excluding LTGs) increased at the highest local densities.

A handful of subsequent studies based on individual clusters also report an increased fraction of slow rotators in dense cluster centres \citep{deugenioetal2013,houghtonetal2013,scottetal2014,fogartyetal2014}.
However, although the fraction of slow rotators was higher in each cluster centre than in the outskirts, the overall fraction of slow rotators in each cluster did not depend on the size of the cluster or on the large-scale density of the cluster's environment.
\citet{jimmyetal2013} searched for signs of recent merging in several brightest cluster galaxies and companions, and find no particular connection between merging signatures and galaxy rotation.
Recently, \citet{olivaetal2017} find only a tentative increase in the fraction of slow rotators with cluster mass for central galaxies.

Most existing studies to date have been limited to a small number of rich environments, or a small volume probing the field.
It is difficult, using these data sets, to decouple the correlated impact of stellar mass and environment on the demographics of slow and fast rotators.
In order to control for stellar mass when investigating the importance of environment, it is necessary to span a full range of environments at the highest masses.

We designed the volume-limited MASSIVE survey to investigate systematically
the most massive galaxies in the northern sky within a distance of 108 Mpc \citep{maetal2014}, targeting 116 galaxies with $M_K \la -25.3$~mag ($M_* \ga 10^{11.5} \; M_\odot$).
This volume is large enough to sample a wide range of environments including several galaxy clusters, many galaxy groups, and  galaxies in the field.
The sample is thus complementary to the \atd\ survey, which includes about twice as many galaxies but from a volume about ten times smaller; it is dominated by the Virgo cluster, and contains only six galaxies more massive than $10^{11.5} \; M_\odot$.
Details of the kinematic analysis of our IFS data were presented in \citet{vealeetal2017}, which focused on the brightest 41 galaxies ($M_K<-25.7$ mag) in the MASSIVE survey.
We have since completed observations and analysis of the larger sample of 75 galaxies with a limiting magnitude of $M_K = -25.5$ mag.

In this paper we perform a detailed analysis of the environments of the entire MASSIVE sample and present measurements of the spin of the 75 MASSIVE galaxies with Mitchell/VIRUS-P IFS data.  
Together with the \atd\ sample of ETGs at lower masses, we investigate the influence of galaxy mass and environment on the spin
of ETGs over a wide range of stellar mass ($8\times 10^{9} \la M_* \la 2\times 10^{12} M_\odot$) and environment.
In particular, the combined MASSIVE and \atd\ sample from the two volume-limited IFS surveys has well-defined stellar mass selection
and is large enough for us to conduct the first analysis of the relationship between spin and environment at fixed $M_*$ for present-day ETGs, and assess how much of the kinematic morphology-density relation is driven by stellar mass.

Since different methods for quantifying environment probe different physical scales, we compare four approaches in this paper: (1) group membership, i.e., if a galaxy is the brightest galaxy or a satellite in a group/cluster, or is relatively isolated; (2) halo mass; (3) smoothed large-scale galaxy density field; and (4) local galaxy density based on the $N$th nearest neighbour.  
The fact that these measures cover differing length scale implies that they
may correlate differently with quantities such as galaxy merger rates, assembly histories, and masses.  

\mautoref{sec:samples} of this paper summarizes the selection and properties of MASSIVE survey galaxies and \atd\ galaxies.
\mautoref{sec:env} presents results for the individual measurements and the statistics of the four environmental quantities.
Technical details of our local density calculation are in \aref{sec:morenu}.
\mautoref{sec:lam} summarizes the kinematic analysis of our IFS data and presents results for galaxy spin versus stellar mass.
\mautoref{sec:lam-env} connects spin to environment, and explores how to decouple those trends from the influence of stellar mass.
\mautoref{sec:summary} discusses implications and conclusions.
Details of the error bars we use to determine whether any trends are significant are presented in \aref{sec:err}. An application of our analysis to the original \atd\ sample densities is presented in \aref{sec:atd}.
We assume $h=0.7$ throughout the paper.


\section{Galaxy Samples}
\label{sec:samples}


\begin{table*}
  \caption{Properties of MASSIVE survey galaxies}
\label{table:main}
\begin{tabular}{lcccc@{\hspace{1em}}c@{\hspace{1em}}cccc@{\hspace{0.5em}}c@{\hspace{0.5em}}ccr}
\hline
Galaxy & R.A. & Dec. & $D$ & $M_K$ & $\log_{10} M_*$ & $\varepsilon$ & $\lambda_e$ & Rot. & env & $\log_{10} M_{\rm halo}$ & Cluster & $(1+\delta_g)$ & $\nu_{10}$ \\
 & [deg] & [deg] & [Mpc] & [mag] & [$M_\odot$] &  &  &  &  & [$M_\odot$] &  &  & [$\overline{\nu}$] \\
(1) & (2) & (3) & (4) & (5) & (6) & (7) & (8) & (9) & (10) & (11) & (12) & (13) & (14) \\
\hline
NGC 0057 & $3.8787$ & $17.3284$ & $76.3$ & $-25.75$ & $11.79$ & $0.17$ & $0.02$ & S & I &  &  & $2.3$ & $4.9$ \\
NGC 0080 & $5.2952$ & $22.3572$ & $81.9$ & $-25.66$ & $11.75$ & $0.09$ & $0.04$ & S & B & $14.1$ &  & $3.0$ & $6600$ \\
NGC 0128 & $7.3128$ & $2.8641$ & $59.3$ & $-25.35$ & $11.61$ & $0.59$ &  &  & I &  &  & $1.4$ & $7.8$ \\
NGC 0227 & $10.6534$ & $-1.5288$ & $75.9$ & $-25.32$ & $11.60$ & $0.25$ &  &  & B & $13.5$ &  & $4.0$ & $4.6$ \\
NGC 0315 & $14.4538$ & $30.3524$ & $70.3$ & $-26.30$ & $12.03$ & $0.28$ & $0.06$ & S & B & $13.5$ &  & $6.0$ & $280$ \\
NGC 0383 & $16.8540$ & $32.4126$ & $71.3$ & $-25.81$ & $11.82$ & $0.14$ & $0.25$ & F & S & $14.4$ &  & $7.2$ & $4400$ \\
NGC 0393 & $17.1540$ & $39.6443$ & $85.7$ & $-25.44$ & $11.65$ & $0.18^*$ &  &  & I &  &  & $1.5$ & $1.4$ \\
NGC 0410 & $17.7453$ & $33.1520$ & $71.3$ & $-25.90$ & $11.86$ & $0.25$ & $0.03$ & S & B & $14.4$ &  & $7.4$ & $3200$ \\
NGC 0467 & $19.7922$ & $3.3008$ & $75.8$ & $-25.40$ & $11.64$ & $0.05$ &  &  & I &  &  & $3.9$ & $18$ \\
PGC 004829 & $20.1287$ & $50.1445$ & $99.0$ & $-25.30$ & $11.59$ & $0.34^*$ &  &  & I &  &  & $2.6$ & $10$ \\
NGC 0499 & $20.7978$ & $33.4601$ & $69.8$ & $-25.50$ & $11.68$ & $0.35$ & $0.06$ & S & S & $14.4$ &  & $7.2$ & $36000$ \\
NGC 0507 & $20.9164$ & $33.2561$ & $69.8$ & $-25.93$ & $11.87$ & $0.09$ & $0.05$ & S & B & $14.4$ &  & $7.2$ & $59000$ \\
NGC 0533 & $21.3808$ & $1.7590$ & $77.9$ & $-26.05$ & $11.92$ & $0.26$ & $0.03$ & S & B & $13.5$ &  & $4.3$ & $13$ \\
NGC 0545 & $21.4963$ & $-1.3402$ & $74.0$ & $-25.83$ & $11.83$ & $0.28$ & $0.13$ & S & B & $14.5$ & A194 & $5.9$ & $14000$ \\
NGC 0547 & $21.5024$ & $-1.3451$ & $74.0$ & $-25.83$ & $11.83$ & $0.14$ & $0.06$ & S & S & $14.5$ & A194 & $5.9$ & $14000$ \\
NGC 0665 & $26.2338$ & $10.4230$ & $74.6$ & $-25.51$ & $11.68$ & $0.24$ & $0.40$ & F & B & $13.7$ &  & $3.0$ & $58$ \\
UGC 01332 & $28.0755$ & $48.0878$ & $99.2$ & $-25.57$ & $11.71$ & $0.30^*$ & $0.04$ & S & B & $13.8$ &  & $3.7$ & $170$ \\
NGC 0708 & $28.1937$ & $36.1518$ & $69.0$ & $-25.65$ & $11.75$ & $0.40^*$ & $0.04$ & S & B & $14.5$ & A262 & $5.8$ & $12000$ \\
UGC 01389 & $28.8778$ & $47.9550$ & $99.2$ & $-25.41$ & $11.64$ & $0.34^*$ &  &  & S & $13.8$ &  & $3.8$ & $150$ \\
NGC 0741 & $29.0874$ & $5.6289$ & $73.9$ & $-26.06$ & $11.93$ & $0.17$ & $0.04$ & S & B & $13.8$ &  & $2.9$ & $130$ \\
NGC 0777 & $30.0622$ & $31.4294$ & $72.2$ & $-25.94$ & $11.87$ & $0.17$ & $0.05$ & S & B & $13.5$ &  & $5.0$ & $78$ \\
NGC 0890 & $35.5042$ & $33.2661$ & $55.6$ & $-25.50$ & $11.68$ & $0.38^*$ & $0.10$ & S & I &  &  & $4.7$ & $1.4$ \\
NGC 0910 & $36.3616$ & $41.8243$ & $79.8$ & $-25.33$ & $11.61$ & $0.16^*$ &  &  & S & $14.8$ & A347 & $6.2$ & $12000$ \\
NGC 0997 & $39.3103$ & $7.3056$ & $90.4$ & $-25.40$ & $11.64$ & $0.13$ &  &  & B & $13.0$ &  & $3.0$ & $26$ \\
NGC 1016 & $39.5815$ & $2.1193$ & $95.2$ & $-26.33$ & $12.05$ & $0.06$ & $0.03$ & S & B & $13.9$ &  & $4.8$ & $56$ \\
NGC 1060 & $40.8127$ & $32.4250$ & $67.4$ & $-26.00$ & $11.90$ & $0.24$ & $0.02$ & S & B & $14.0$ &  & $3.9$ & $2100$ \\
NGC 1066 & $40.9579$ & $32.4749$ & $67.4$ & $-25.31$ & $11.60$ & $0.16$ &  &  & S & $14.0$ &  & $3.9$ & $2200$ \\
NGC 1132 & $43.2159$ & $-1.2747$ & $97.6$ & $-25.70$ & $11.77$ & $0.37$ & $0.06$ & S & B & $13.6$ &  & $3.4$ & $8.3$ \\
NGC 1129 & $43.6141$ & $41.5796$ & $73.9$ & $-26.14$ & $11.96$ & $0.15^\dagger$ & $0.12$ & S & B & $14.8$ &  & $10.7$ & $16000$ \\
NGC 1167 & $45.4265$ & $35.2056$ & $70.2$ & $-25.64$ & $11.74$ & $0.17$ & $0.43$ & F & B & $13.1$ &  & $5.0$ & $15$ \\
NGC 1226 & $47.7723$ & $35.3868$ & $85.7$ & $-25.51$ & $11.68$ & $0.18^*$ & $0.03$ & S & B & $13.2$ &  & $3.5$ & $3.1$ \\
IC0 310 & $49.1792$ & $41.3248$ & $77.5$ & $-25.35$ & $11.61$ & $0.06$ &  &  & S & $14.8$ & Perseus & $13.2$ & $15000$ \\
NGC 1272 & $49.8387$ & $41.4906$ & $77.5$ & $-25.80$ & $11.81$ & $0.07$ & $0.02$ & S & S & $14.8$ & Perseus & $13.5$ & $400000$ \\
UGC 02783 & $53.5766$ & $39.3568$ & $85.8$ & $-25.44$ & $11.65$ & $0.11$ &  &  & B & $12.6$ &  & $6.3$ & $17$ \\
NGC 1453 & $56.6136$ & $-3.9688$ & $56.4$ & $-25.67$ & $11.75$ & $0.14^*$ & $0.20$ & F & B & $13.9$ &  & $2.3$ & $89$ \\
NGC 1497 & $60.5283$ & $23.1329$ & $87.8$ & $-25.31$ & $11.60$ & $0.40^*$ &  &  & I &  &  & $2.7$ & $89$ \\
NGC 1600 & $67.9161$ & $-5.0861$ & $63.8$ & $-25.99$ & $11.90$ & $0.26^*$ & $0.03$ & S & B & $14.2$ &  & $6.0$ & $1200$ \\
NGC 1573 & $68.7666$ & $73.2624$ & $65.0$ & $-25.55$ & $11.70$ & $0.34^*$ & $0.04$ & S & B & $14.1$ &  & $4.1$ & $590$ \\
NGC 1684 & $73.1298$ & $-3.1061$ & $63.5$ & $-25.34$ & $11.61$ & $0.24^*$ &  &  & B & $13.7$ &  & $6.2$ & $1600$ \\
NGC 1700 & $74.2347$ & $-4.8658$ & $54.4$ & $-25.60$ & $11.72$ & $0.28^*$ & $0.20$ & F & B & $12.7$ &  & $3.5$ & $23$ \\
NGC 2208 & $95.6444$ & $51.9095$ & $84.1$ & $-25.63$ & $11.74$ & $0.32^*$ & $0.06$ & S & I &  &  & $2.8$ & $7.2$ \\
NGC 2256 & $101.8082$ & $74.2365$ & $79.4$ & $-25.87$ & $11.84$ & $0.20^*$ & $0.02$ & S & B & $13.7$ &  & $2.7$ & $21$ \\
NGC 2274 & $101.8224$ & $33.5672$ & $73.8$ & $-25.69$ & $11.76$ & $0.10^*$ & $0.07$ & S & B & $13.3$ &  & $3.1$ & $110$ \\
NGC 2258 & $101.9425$ & $74.4818$ & $59.0$ & $-25.66$ & $11.75$ & $0.24^*$ & $0.04$ & S & B & $12.2$ &  & $3.8$ & $9.8$ \\
NGC 2320 & $106.4251$ & $50.5811$ & $89.4$ & $-25.93$ & $11.87$ & $0.30^*$ & $0.23$ & F & B & $14.2$ &  & $7.9$ & $660$ \\
UGC 03683 & $107.0582$ & $46.1159$ & $85.1$ & $-25.52$ & $11.69$ & $0.26^*$ & $0.09$ & S & B & $13.6$ &  & $5.8$ & $27$ \\
NGC 2332 & $107.3924$ & $50.1823$ & $89.4$ & $-25.39$ & $11.63$ & $0.34^*$ &  &  & S & $14.2$ &  & $7.8$ & $1500$ \\
NGC 2340 & $107.7950$ & $50.1747$ & $89.4$ & $-25.90$ & $11.86$ & $0.44^*$ & $0.03$ & S & S & $14.2$ &  & $7.8$ & $1300$ \\
UGC 03894 & $113.2695$ & $65.0791$ & $97.2$ & $-25.58$ & $11.72$ & $0.10$ & $0.12$ & F & B & $13.7$ &  & $1.5$ & $1.5$ \\
NGC 2418 & $114.1563$ & $17.8839$ & $74.1$ & $-25.42$ & $11.64$ & $0.20$ &  &  & I &  &  & $2.2$ & $1.4$ \\
NGC 2456 & $118.5444$ & $55.4953$ & $107.3$ & $-25.33$ & $11.61$ & $0.24^*$ &  &  & I &  &  & $2.4$ & $3.7$ \\
NGC 2492 & $119.8738$ & $27.0264$ & $97.8$ & $-25.36$ & $11.62$ & $0.19$ &  &  & B & $13.0$ &  & $1.1$ & $0.8$ \\
NGC 2513 & $120.6028$ & $9.4136$ & $70.8$ & $-25.52$ & $11.69$ & $0.20$ & $0.10$ & S & B & $13.6$ &  & $2.3$ & $5.2$ \\
NGC 2672 & $132.3412$ & $19.0750$ & $61.5$ & $-25.60$ & $11.72$ & $0.14$ & $0.10$ & S & B & $13.0$ &  & $1.3$ & $1.3$ \\
NGC 2693 & $134.2469$ & $51.3474$ & $74.4$ & $-25.76$ & $11.79$ & $0.25$ & $0.29$ & F & I &  &  & $1.7$ & $6.9$ \\
NGC 2783 & $138.4145$ & $29.9929$ & $101.4$ & $-25.72$ & $11.78$ & $0.39$ & $0.04$ & S & B & $12.8$ &  & $3.2$ & $4.7$ \\
NGC 2832 & $139.9453$ & $33.7498$ & $105.2$ & $-26.42$ & $12.08$ & $0.31$ & $0.07$ & S & B & $13.7$ & A779 & $4.0$ & $7.9$ \\
NGC 2892 & $143.2205$ & $67.6174$ & $101.1$ & $-25.70$ & $11.77$ & $0.06$ & $0.05$ & S & I &  &  & $2.2$ & $2.3$ \\
NGC 2918 & $143.9334$ & $31.7054$ & $102.3$ & $-25.49$ & $11.68$ & $0.17$ &  &  & I &  &  & $3.0$ & $2.5$ \\
NGC 3158 & $153.4605$ & $38.7649$ & $103.4$ & $-26.28$ & $12.02$ & $0.18$ & $0.26$ & F & B & $13.3$ &  & $2.7$ & $9.8$ \\
NGC 3209 & $155.1601$ & $25.5050$ & $94.6$ & $-25.55$ & $11.70$ & $0.27$ & $0.04$ & S & B & $11.8$ &  & $2.4$ & $2.8$ \\
NGC 3332 & $160.1182$ & $9.1825$ & $89.1$ & $-25.38$ & $11.63$ & $0.16$ &  &  & I &  &  & $1.0$ & $0.6$ \\
\end{tabular}
\end{table*}
\begin{table*}
\contcaption{}
\label{table:maincontinued}
\begin{tabular}{lcccc@{\hspace{1em}}c@{\hspace{1em}}cccc@{\hspace{0.5em}}c@{\hspace{0.5em}}ccr}
\hline
Galaxy & R.A. & Dec. & $D$ & $M_K$ & $\log_{10} M_*$ & $\varepsilon$ & $\lambda_e$ & Rot. & env & $\log_{10} M_{\rm halo}$ & Cluster & $(1+\delta_g)$ & $\nu_{10}$ \\
 & [deg] & [deg] & [Mpc] & [mag] & [$M_\odot$] &  &  &  &  & [$M_\odot$] &  &  & [$\overline{\nu}$] \\
(1) & (2) & (3) & (4) & (5) & (6) & (7) & (8) & (9) & (10) & (11) & (12) & (13) & (14) \\
\hline
NGC 3343 & $161.5432$ & $73.3531$ & $93.8$ & $-25.33$ & $11.61$ & $0.32^*$ &  &  & I &  &  & $2.0$ & $16$ \\
NGC 3462 & $163.8378$ & $7.6967$ & $99.2$ & $-25.62$ & $11.73$ & $0.26$ & $0.09$ & S & I &  &  & $2.2$ & $2.5$ \\
NGC 3562 & $168.2445$ & $72.8793$ & $101.0$ & $-25.65$ & $11.75$ & $0.16^*$ & $0.04$ & S & B & $13.5$ &  & $2.2$ & $8.5$ \\
NGC 3615 & $169.5277$ & $23.3973$ & $101.2$ & $-25.58$ & $11.72$ & $0.38$ & $0.40$ & F & B & $13.6$ &  & $3.1$ & $5.2$ \\
NGC 3805 & $175.1736$ & $20.3430$ & $99.4$ & $-25.69$ & $11.76$ & $0.36$ & $0.50$ & F & S & $14.8$ & A1367 & $5.6$ & $440$ \\
NGC 3816 & $175.4502$ & $20.1036$ & $99.4$ & $-25.40$ & $11.64$ & $0.31$ &  &  & S & $14.8$ & A1367 & $5.8$ & $1900$ \\
NGC 3842 & $176.0090$ & $19.9498$ & $99.4$ & $-25.91$ & $11.86$ & $0.22$ & $0.04$ & S & B & $14.8$ & A1367 & $5.9$ & $19000$ \\
NGC 3862 & $176.2708$ & $19.6063$ & $99.4$ & $-25.50$ & $11.68$ & $0.06$ & $0.06$ & S & S & $14.8$ & A1367 & $5.9$ & $18000$ \\
NGC 3937 & $178.1776$ & $20.6313$ & $101.2$ & $-25.62$ & $11.73$ & $0.20$ & $0.07$ & S & B & $14.2$ &  & $5.9$ & $71$ \\
NGC 4055 & $181.0059$ & $20.2323$ & $107.2$ & $-25.40$ & $11.64$ & $0.18$ &  &  & S & $14.3$ &  & $7.1$ & $2300$ \\
NGC 4065 & $181.0257$ & $20.2351$ & $107.2$ & $-25.47$ & $11.67$ & $0.22$ &  &  & B & $14.3$ &  & $7.1$ & $2500$ \\
NGC 4066 & $181.0392$ & $20.3479$ & $107.2$ & $-25.35$ & $11.61$ & $0.06$ &  &  & S & $14.3$ &  & $7.1$ & $4200$ \\
NGC 4059 & $181.0471$ & $20.4098$ & $107.2$ & $-25.41$ & $11.64$ & $0.06$ &  &  & S & $14.3$ &  & $7.1$ & $4900$ \\
NGC 4073 & $181.1128$ & $1.8960$ & $91.5$ & $-26.33$ & $12.05$ & $0.32$ & $0.02$ & S & B & $13.9$ &  & $4.4$ & $89$ \\
NGC 4213 & $183.9064$ & $23.9819$ & $101.6$ & $-25.44$ & $11.65$ & $0.17$ &  &  & B & $13.4$ &  & $4.7$ & $16$ \\
NGC 4472 & $187.4450$ & $8.0004$ & $16.7$ & $-25.72$ & $11.78$ & $0.17^\dagger$ & $0.20$ & F & B & $14.7$ & Virgo & $8.9$ & $1900$ \\
NGC 4486 & $187.7059$ & $12.3911$ & $16.7$ & $-25.31$ & $11.60$ & $0.10$ &  &  & S & $14.7$ & Virgo & $9.1$ & $14000$ \\
NGC 4555 & $188.9216$ & $26.5230$ & $103.6$ & $-25.92$ & $11.86$ & $0.20$ & $0.12$ & S & I &  &  & $5.9$ & $6.3$ \\
NGC 4649 & $190.9167$ & $11.5526$ & $16.5$ & $-25.36$ & $11.62$ & $0.20$ &  &  & S & $14.7$ & Virgo & $9.1$ & $2600$ \\
NGC 4816 & $194.0506$ & $27.7455$ & $102.0$ & $-25.33$ & $11.61$ & $0.20$ &  &  & S & $15.3$ & Coma & $13.2$ & $1900$ \\
NGC 4839 & $194.3515$ & $27.4977$ & $102.0$ & $-25.85$ & $11.83$ & $0.35$ & $0.05$ & S & S & $15.3$ & Coma & $13.2$ & $2600$ \\
NGC 4874 & $194.8988$ & $27.9594$ & $102.0$ & $-26.18$ & $11.98$ & $0.09$ & $0.07$ & S & S & $15.3$ & Coma & $13.2$ & $24000$ \\
NGC 4889 & $195.0338$ & $27.9770$ & $102.0$ & $-26.64$ & $12.18$ & $0.36$ & $0.03$ & S & B & $15.3$ & Coma & $13.2$ & $19000$ \\
NGC 4914 & $195.1789$ & $37.3153$ & $74.5$ & $-25.72$ & $11.78$ & $0.39$ & $0.05$ & S & I &  &  & $1.1$ & $1.2$ \\
NGC 5129 & $201.0417$ & $13.9765$ & $107.5$ & $-25.92$ & $11.86$ & $0.37$ & $0.40$ & F & I &  &  & $4.3$ & $4.9$ \\
NGC 5208 & $203.1163$ & $7.3166$ & $105.0$ & $-25.61$ & $11.73$ & $0.63$ & $0.61$ & F & B & $13.0$ &  & $5.0$ & $16$ \\
PGC 047776 & $203.4770$ & $3.2836$ & $103.8$ & $-25.36$ & $11.62$ & $0.18$ &  &  & B & $14.1$ &  & $4.0$ & $17$ \\
NGC 5252 & $204.5661$ & $4.5426$ & $103.8$ & $-25.32$ & $11.60$ & $0.52$ &  &  & S & $14.1$ &  & $4.9$ & $52$ \\
NGC 5322 & $207.3133$ & $60.1904$ & $34.2$ & $-25.51$ & $11.68$ & $0.33$ & $0.05$ & S & B & $13.7$ &  & $2.5$ & $21$ \\
NGC 5353 & $208.3613$ & $40.2831$ & $41.1$ & $-25.45$ & $11.66$ & $0.56$ &  &  & B & $13.6$ &  & $2.6$ & $63$ \\
NGC 5490 & $212.4888$ & $17.5455$ & $78.6$ & $-25.57$ & $11.71$ & $0.20$ & $0.14$ & S & I &  &  & $2.1$ & $9.8$ \\
NGC 5557 & $214.6071$ & $36.4936$ & $51.0$ & $-25.46$ & $11.66$ & $0.17$ &  &  & B & $13.3$ &  & $2.6$ & $8.5$ \\
IC1 143 & $232.7345$ & $82.4558$ & $97.3$ & $-25.45$ & $11.66$ & $0.14^*$ &  &  & B & $13.0$ &  & $2.0$ & $13$ \\
UGC 10097 & $238.9303$ & $47.8673$ & $91.5$ & $-25.43$ & $11.65$ & $0.23$ &  &  & B & $12.7$ &  & $1.5$ & $5.0$ \\
NGC 6223 & $250.7679$ & $61.5789$ & $86.7$ & $-25.59$ & $11.72$ & $0.20^*$ & $0.32$ & F & B & $13.5$ &  & $1.5$ & $6.2$ \\
NGC 6364 & $261.1139$ & $29.3902$ & $105.3$ & $-25.38$ & $11.63$ & $0.15$ &  &  & I &  &  & $0.8$ & $0.5$ \\
NGC 6375 & $262.3411$ & $16.2067$ & $95.8$ & $-25.53$ & $11.69$ & $0.10^*$ & $0.24$ & F & I &  &  & $1.2$ & $1.5$ \\
UGC 10918 & $264.3892$ & $11.1217$ & $100.2$ & $-25.75$ & $11.79$ & $0.14^*$ & $0.03$ & S & I &  &  & $1.8$ & $4.8$ \\
NGC 6442 & $266.7139$ & $20.7611$ & $98.0$ & $-25.40$ & $11.64$ & $0.12^*$ &  &  & I &  &  & $1.1$ & $3.0$ \\
NGC 6482 & $267.9534$ & $23.0719$ & $61.4$ & $-25.60$ & $11.72$ & $0.36^*$ & $0.14$ & S & B & $13.1$ &  & $1.6$ & $1.1$ \\
NGC 6575 & $272.7395$ & $31.1162$ & $106.0$ & $-25.58$ & $11.72$ & $0.28^*$ & $0.12$ & S & I &  &  & $2.1$ & $5.0$ \\
NGC 7052 & $319.6377$ & $26.4469$ & $69.3$ & $-25.67$ & $11.75$ & $0.50^*$ & $0.15$ & S & I &  &  & $1.3$ & $0.8$ \\
NGC 7242 & $333.9146$ & $37.2987$ & $84.4$ & $-26.34$ & $12.05$ & $0.28^*$ & $0.04$ & S & B & $14.0$ &  & $6.3$ & $2800$ \\
NGC 7265 & $335.6145$ & $36.2098$ & $82.8$ & $-25.93$ & $11.87$ & $0.22^*$ & $0.04$ & S & B & $14.7$ &  & $6.9$ & $5200$ \\
NGC 7274 & $336.0462$ & $36.1259$ & $82.8$ & $-25.39$ & $11.63$ & $0.06^*$ &  &  & S & $14.7$ &  & $6.9$ & $3200$ \\
NGC 7386 & $342.5089$ & $11.6987$ & $99.1$ & $-25.58$ & $11.72$ & $0.28$ & $0.07$ & S & B & $13.9$ &  & $2.6$ & $3.2$ \\
NGC 7426 & $344.0119$ & $36.3614$ & $80.0$ & $-25.74$ & $11.79$ & $0.34^*$ & $0.56$ & F & B & $13.8$ &  & $3.8$ & $8.5$ \\
NGC 7436 & $344.4897$ & $26.1500$ & $106.6$ & $-26.16$ & $11.97$ & $0.12$ & $0.09$ & S & B & $14.4$ &  & $4.1$ & $100$ \\
NGC 7550 & $348.8170$ & $18.9614$ & $72.7$ & $-25.43$ & $11.65$ & $0.07$ &  &  & B & $11.9$ &  & $0.9$ & $1.0$ \\
NGC 7556 & $348.9353$ & $-2.3815$ & $103.0$ & $-25.83$ & $11.83$ & $0.25$ & $0.05$ & S & B & $14.0$ &  & $2.0$ & $17$ \\
NGC 7618 & $349.9468$ & $42.8526$ & $76.3$ & $-25.44$ & $11.65$ & $0.28^*$ &  &  & B & $13.7$ &  & $3.2$ & $250$ \\
NGC 7619 & $350.0605$ & $8.2063$ & $54.0$ & $-25.65$ & $11.75$ & $0.23$ & $0.12$ & S & B & $14.0$ &  & $1.5$ & $22$ \\
NGC 7626 & $350.1772$ & $8.2170$ & $54.0$ & $-25.65$ & $11.75$ & $0.14$ & $0.03$ & S & S & $14.0$ &  & $1.5$ & $21$ \\
\end{tabular}

  Column notes:
  (1) Galaxy name (in order of increasing right ascension).
  (2), (3) Right Ascension, Declination in degrees.
  (4) Distance according to \citet{maetal2014}.
  (5) Extinction-corrected total absolute $K$-band magnitude.
  (6) Stellar mass estimated from $M_K$.
  (7) Ellipticity from NSA where available, from 2MASS otherwise (asterisks). $^\dagger$ NGC 1129 and NGC 4472 are from our CFHT data and \citet{emsellemetal2011} respectively; see \citet{vealeetal2017} for details.
  (8) Proxy for the spin parameter within the effective radius.
  (9) Slow or fast rotator classification. Most galaxies are slow rotators (``S''), with few fast rotators (``F''). See \mautoref{sec:lam-def} for definitions.
  (10) Group membership according to the HDC catalogue. Most galaxies are BGG (``B''), with few satellite (``S'') or isolated (``I'') galaxies.
  (11) Halo mass according to the HDC catalogue, or from updated literature sources (see text) for Virgo, Coma, and Perseus.
  (12) Membership in Virgo, Coma, Perseus, or Abell clusters.
  (13) Large-scale galaxy overdensity from the 2M++ catalogue (\mautoref{sec:delta}).
  (14) Local density in units of the mean $K$-band luminosity density $\overline{\nu} \sim 2.8 \times 10^8\; {\rm L}_\odot \; {\rm Mpc}^{-3}$ (\mautoref{sec:nu}).
\end{table*}


The MASSIVE survey consists of a volume-limited sample of early-type galaxies, targeting all 116 galaxies\footnote{The actual count is 115 galaxies after we remove NGC 7681, as discussed in \citet{vealeetal2017}, which our data showed to be a close pair of bulges.
We likewise exclude NGC 7681 from this paper.}
with $K$-band magnitudes $M_K$ brighter than $-25.3$ mag (i.e. stellar masses $M^* \gtrsim 10^{11.5} M_\odot$) and distances within $D < 108$ Mpc, in the northern hemisphere and away from the galactic plane \citep{maetal2014}.
The galaxies were selected from the Extended Source Catalogue \citep[XSC;][]{jarrettetal2000} of the Two Micron All Sky Survey \citep[2MASS;][]{skrutskieetal2006}.
Distances are taken from the surface-brightness fluctuation method \citep{blakesleeetal2009,blakesleeetal2010,blakeslee2013} for galaxies in Virgo and Coma, from the High Density Contrast (HDC) group catalogue \citep[][see also \mautoref{sec:bgg}]{crooketal2007} for other galaxies if available, and from the flow model of \citet{mouldetal2000} otherwise.
We use $M_K$ as a proxy for the stellar mass, estimated using equation 2 of \citet{maetal2014}.
That equation was taken from equation 2 of \citet{cappellari2013}, a fit to stellar masses derived from dynamical modeling.

We have completed observations of the ``priority sample'' of the MASSIVE survey, which consists of the 75 galaxies with $M_K < -25.5$ mag ($M_* \ga 10^{11.7} \; M_\odot$).
The observations were performed using the Mitchell/VIRUS-P IFS at the McDonald Observatory \citep{hilletal2008}, which has a large 107\arcsec$\times$107\arcsec\ field of view and consists of 246 evenly-spaced 4\arcsec-diameter fibres with a one-third filling factor.
We observed each galaxy with three dither positions of equal exposure time to obtain contiguous coverage of the field of view.
The spectral range spanned 3650 \AA\ to 5850 \AA, covering the Ca H+K region, the G-band region, H$\beta$, the Mgb region, and many Fe absorption features,
with $\sim 2000$ pixels (log-spaced in wavelength) and an instrumental resolution of $\sim 4.5$~\AA\ full width at half-maximum.

We spatially bin our IFS spectra to obtain a mean signal-to-noise ratio (S/N) of at least 20 (per pixel) for each spectrum, folding across the major axis to combine symmetrical bins and obtain our minimum S/N with the smallest possible bin size.
To obtain the stellar line-of-sight velocity distribution (LOSVD) for each spectrum, we use the penalized pixel-fitting (pPXF) method of \citet{cappellariemsellem2004} and parametrize the LOSVD as a Gauss-Hermite series up to order 6.
For each spectrum we thus obtain velocity $V$, dispersion $\sigma$, and higher order moments $h_3, h_4, h_5$ and $h_6$.
See \citet{vealeetal2017} for a more detailed description of the analysis.

There are two main sources of uncertainty on $M_K$.
First, the $K$-band magnitudes of 2MASS are likely underestimated somewhat due to the shallowness of the survey \citep{laueretal2007,schombertsmith2012}.
Second, the choice of distance estimate and extinction (see \citealt{maetal2014} for details) can also impact $M_K$ and $M_*$; based on the galaxies in common between MASSIVE and \atd, typical differences in $M_K$ due to different estimates is around 0.1~mag, and up to nearly 0.5~mag for extreme cases.
These are both larger than the uncertainty due to formal errors in $K$-band magnitude from 2MASS, which are generally less than $0.03$~mag.
Together, combined with the 0.14 dex scatter in the $M_*$-$M_K$ relation \citep{cappellari2013}, these correspond to an underestimation of $M_*$ of up to $\sim0.3$~dex, and an uncertainty of $\sim 0.2$~dex.

Where available, additional photometric data is taken from the NASA-Sloan Atlas (NSA, http://www.nsatlas.org) based on the SDSS DR8 catalogue \citep{yorketal2000,aiharaetal2011}; otherwise the 2MASS values are used.
The effective radius $R_e$ from these sources, like $M_K$, may be underestimated due to the shallowness of the surveys.
We have scaled the 2MASS $R_e$ to be comparable to the NSA values \citep[see][for details]{maetal2014} to partly account for the 2MASS underestimation.
Comparing ellipticity $\varepsilon$ and position angle (PA) of galaxies with both NSA and 2MASS data gives an estimate of uncertainties in those quantities of $\sim 0.05$ and $\sim 15$~degrees, respectively.
Very round galaxies may have much higher uncertainties in PA; we discuss in \autoref{sec:lam-def} how this impacts our kinematic analysis, and why we do not expect these uncertainties (or the underestimation of $R_e$) to impact our results.
Results from our deep CFHT photometry will be reported in upcoming MASSIVE papers.

As a comparison sample to MASSIVE, we examine the 260 nearby galaxies in the \atd\ survey (also selected from the 2MASS XSC), which targets all early-type galaxies with $M_K < -21.5$ mag ($M_* \ga 10^{9.9} \; M_\odot$) and located within $D < 42$ Mpc, also in the northern hemisphere and excluding the galactic plane \citep{cappellarietal2011a}.
Due to the larger volume and brighter $M_K$ cutoff of the MASSIVE survey, only 6 \atd\ galaxies overlap with MASSIVE.  The two surveys therefore target complementary parameter space in stellar mass and volume. 
Of the 6 common galaxies, NGC 4472 and NGC 5322 are in the priority sample ($M_K < -25.5$ mag) presented in this paper, while the remaining 4 (NGC 4486, 4649, 5353, 5557) are fainter than this limit.  
Our kinematic measurements of $V$, $\sigma$, $h_3$ and $h_4$ agree well with \atd\ for the inner $\sim 25$\arcsec\ region of each galaxy covered by \atd, but our data extend to at least $\sim 60$\arcsec\ in radius (see Figure~B1 of \citealt{vealeetal2017}, where we compare kinematics from both surveys for NGC 4472, 5322, and 5557).
We use our wide-field kinematic results for NGC 4472 and NGC 5322 below and remove these two galaxies from the \atd\ sample in our plots, to avoid double-counting and create a clean separation in $M_K$ between the two samples.

We note that $\sim 20\%$ of galaxies in the \atd\ sample are in the Virgo cluster.
These galaxies are powerful probes of the intra-cluster environments within Virgo, but they probe only a single cluster.
By contrast, no MASSIVE galaxy resides in a galaxy group or cluster (as defined in \mautoref{sec:bgg}) containing more than three other MASSIVE galaxies.
The MASSIVE sample therefore tends to probe distinct group/cluster environments.
When appropriate, we will denote Virgo galaxies in the \atd\ sample with distinct symbols below so that the rest of \atd\ sample can be compared with the MASSIVE sample more fairly.

\section{Galaxy Environments}
\label{sec:env}

\begin{figure*}
\begin{center}
\includegraphics[page=1]{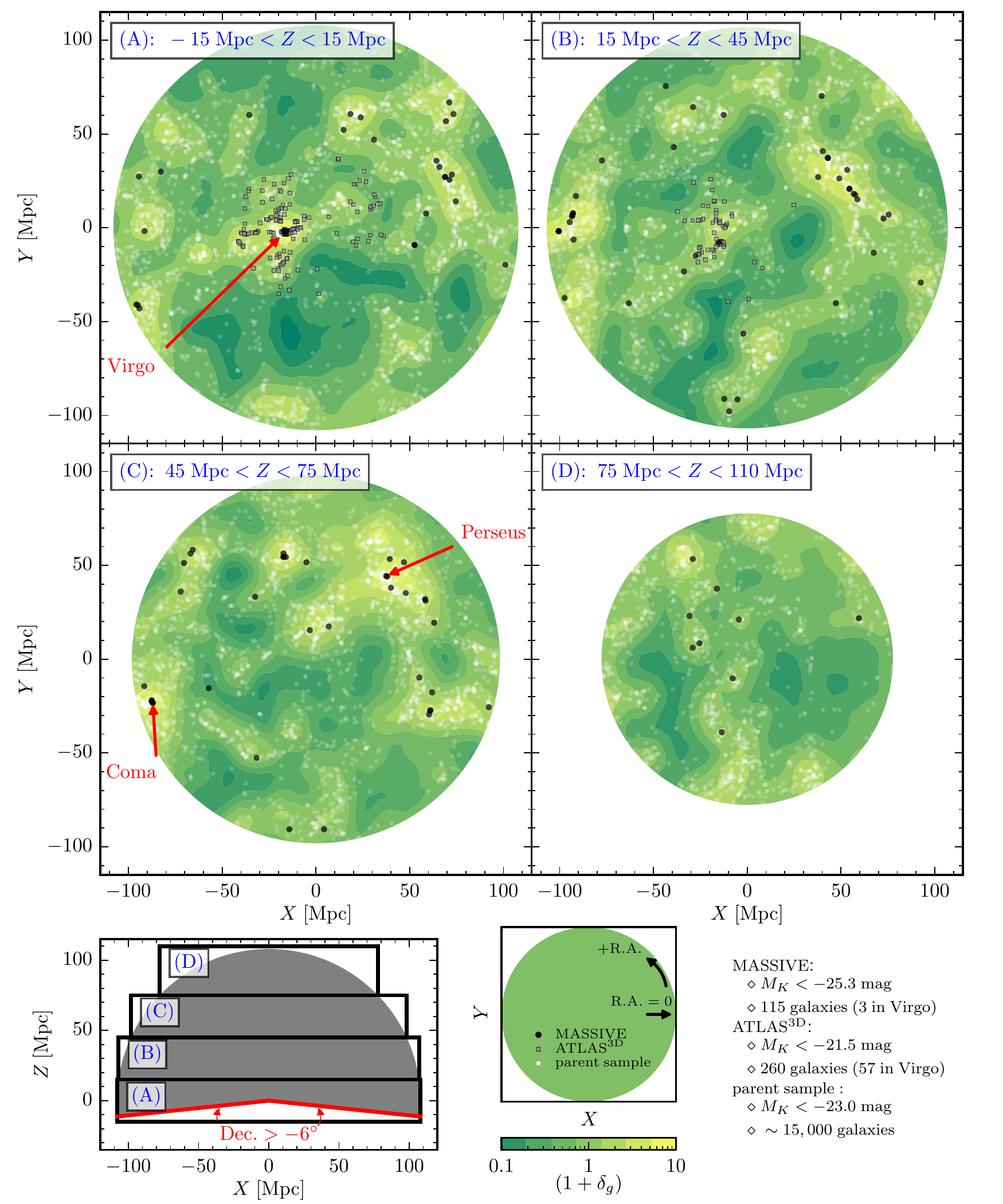}
\end{center}
\caption{Galaxy distribution in the MASSIVE survey volume, where the $X$-$Y$ plane is the earth equatorial plane, in four vertical slices.
Galaxies in the MASSIVE survey (black circles) span a volume more than 10 times that of the \atd\ galaxies (grey squares).
Contour map colours (yellow to dark green) show the large-scale density $(1+\delta_g)$ from the 2M++ Redshift Catalogue, averaged over $Z$ within the slice at each pixel. 
These contours trace closely the galaxy locations of the parent sample selected from 2MRS for purposes of calculating local density $\nu_{10}$ (see \aref{sec:morenu} for details) which are shown as transparent white points (areas of many overlapping galaxies are brighter).
With a cut of $M_K < -23.0$~mag, 2MRS (and hence our parent sample) is nearly complete out to our maximum distance of $D = 108$~Mpc (see \aref{sec:morenu}). 
}
\label{fig:bigmap}
\end{figure*}

In this paper we use four different measures to quantify galaxy environments and to investigate the connection between galaxy environments and stellar kinematics for the galaxies in the MASSIVE and \atd\ surveys: (1) group membership from the group catalogues of \citet{crooketal2007} constructed from the 2MASS Redshift Survey \citep[2MRS;][]{huchraetal2012}; (2) halo mass from the same group catalogues, available for galaxies in a group with 3 or more members; (3) a smoothed large-scale density field from \citet{carricketal2015} based on the 2M++ Redshift Catalogue of \citet{lavauxhudson2011}; and (4) a local galaxy luminosity density calculated within the volume to the 10th neighbour, similar to $\nu_{10}$ in \citet{cappellarietal2011b}.
We discuss the differences and caveats of the four methods in the subsections below.
See also, e.g., \citet{muldrewetal2012} for a comprehensive study of different definitions of galaxy environments,
and \citet{carolloetal2013} for a discussion of how a similar set of environment measures connects to other galaxy properties such as size, color, and star formation rate.
Each measure of environment is tabulated in \mautoref{table:main}.
An overview of the MASSIVE volume is shown in \mautoref{fig:bigmap}.

\subsection{Group Membership}
\label{sec:bgg}

\begin{figure}
\begin{center}
\includegraphics[page=2]{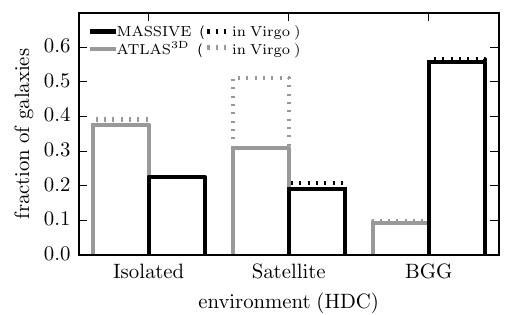}
\end{center}
\caption{Distribution of MASSIVE (black) and \atd\ (grey) galaxies in three environment types based on the 2MRS HDC group catalogue.
The galaxy fractions are computed within each survey: 56\% of MASSIVE galaxies are BGGs, whereas only 10\% of \atd\ galaxies are BGGs.
Virgo galaxies (dotted lines) are stacked above non-Virgo galaxies. About 20\% of \atd\ satellite galaxies are in the Virgo cluster, and some Virgo galaxies are classified as isolated due to different definitions of the cluster boundaries between \atd\ and the 2MRS HDC group catalogue.
  The small fraction of BGGs in \atd\ is expected due to the inclusion of galaxies as faint as $-21.5$ mag, especially the many Virgo galaxies, but may be made even smaller by incompleteness in the HDC catalogue (see text).
}
\label{fig:bgg}
\end{figure}

\citet{crooketal2007,crooketal2008} published redshift-limited catalogues of galaxy groups based on a 2MRS sample complete to an apparent magnitude (corrected for extinction) of $K<11.25$ mag \citep{huchraetal2005a,huchraetal2005b}.
This limiting magnitude corresponds to an absolute magnitude of approximately $M_K < -23.9$ mag at our maximum distance of 108 Mpc, and approximately $M_K < -21.9$ mag at the maximum \atd\ distance of 42 Mpc.
The group catalogues thus cover both MASSIVE ($M_K < -25.3$ mag) and \atd\ galaxies ($M_K < -21.5$ mag), with only two \atd\ galaxies (PGC 029321 and UGC 05408) falling outside the magnitude cut.

\citet{crooketal2007} applied the Friends-of-Friends (FOF) algorithm with two sets of linking parameters to create two group catalogues of differing density contrasts.
The High Density Contrast (HDC) catalogue used a linking length of 350 \kms\ along the line of sight and 0.89 Mpc in the transverse direction,
corresponding to a density contrast of $\delta \rho / \rho \gsim 80$.
The Low Density Contrast (LDC) catalogue used
larger linking lengths of 399 \kms\ and 1.63 Mpc
for a density contrast of $\delta \rho / \rho \gsim 12$.
In \citet{maetal2014} we compared and discussed the agreement between the group assignments of the HDC and the 2M++ redshift catalogue of \citet{lavauxhudson2011}.

We classify MASSIVE and \atd\ galaxies into three types according to their group membership in the HDC catalogue: (1) ``brightest group galaxy'' (BGG) that belongs to a group and is the most luminous galaxy in the group; (2) ``satellite'' that belongs to a group but is not the BGG; and (3) ``isolated'' galaxy that does not belong to a group of at least 3 members in the catalogue.
We make no attempt to determine whether the BGG of a group is also the {\em central} galaxy of the group, which it may not be \citep[e.g.][]{skibbaetal2011}; see \citet{olivaetal2017} for a discussion of rotation in central galaxies.

\autoref{fig:bgg} shows that among the 116 MASSIVE galaxies, 56\% are BGGs, 21\% are satellites in groups, and 23\% are ``isolated'', whereas in \atd, most of the galaxies are either satellites (51\%) or isolated (39\%), with only 10\% being BGGs.
The much lower percentage of BGGs in \atd\ than MASSIVE is largely a result of the smaller survey volume and lower mass limit of the \atd\ survey. It may also be further suppressed by the relative incompleteness of the HDC catalogue near the edges of the \atd\ volume.
While only two \atd\ galaxies are outside the magnitude cut of \citet{crooketal2007} entirely, group membership status also depends on whether nearby galaxies are inside or outside this cut.
For example, the same galaxy might be classified as BGG of a group with 4 members at a close distance, but classified as isolated at a farther distance due to all 3 satellite galaxies falling outside the magnitude cut.
Thus, \atd\ galaxies near the edge of the sample and near the completeness limit of the \citet{crooketal2007} catalogues may be biased towards classification as isolated.

In principle all MASSIVE galaxies beyond the \atd\ volume are also subject to this relative bias towards being classified as isolated, but this will only occur if the (unidentified) rank 2 galaxy is below the $K=11.25$ magnitude cut ($M_K=-23.9$~mag at $D=108$~Mpc, compared to $M_K < -25.3$ for MASSIVE galaxies).
Most groups have a smaller gap between BGG and rank 2 galaxies than this, so we proceed with the classifications as they are.
We are consistent with high-redshift studies showing progenitors of our very massive galaxies in isolated environments \citep{vulcanietal2016}.

We will use both HDC and LDC catalogues to provide group distances in \mautoref{sec:nu} and \aref{sec:morenu}, instead of the improved distances calculated in \citet{maetal2014}, for purposes of calculating local density $\nu_{10}$; see those sections for details.

\subsection{Halo Mass}
\label{sec:mh}

\begin{figure*}
\begin{center}
\includegraphics[page=1]{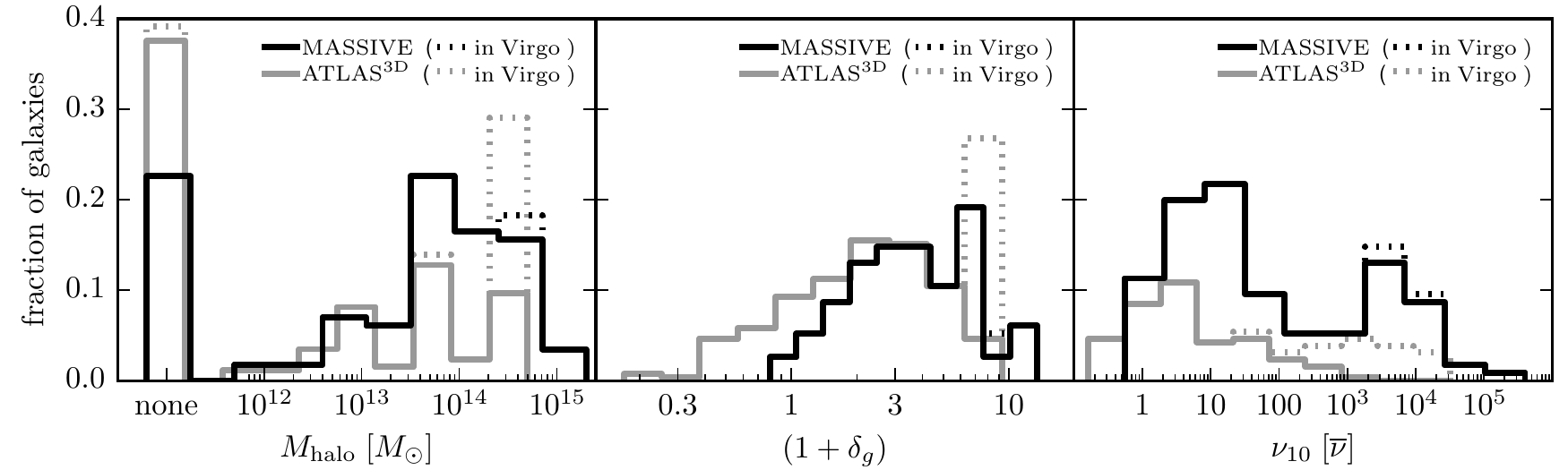}
\end{center}
\caption{Distribution of MASSIVE (black) and \atd\ galaxies (grey) in bins of halo mass $\mh$ (left panel), large scale density $(1+\delta_g)$ (middle panel), and local density $\nu_{10}$ (right panel).
Galaxies with fewer than 3 group members in the 2MRS HDC catalogue do not have $\mh$ measurements; these "isolated" galaxies are placed in the bin labeled "none" in the left panel. 
In each panel, Virgo galaxies (dotted lines) are stacked above non-Virgo galaxies.
We use the \atd\ definition of Virgo (within a sphere of $R=3.5$~Mpc), but some ``non-Virgo'' galaxies are also in the same group as defined by the HDC catalogue, so less than half of the ``non-Virgo'' galaxies in the highest $\mh$ bin are actually in a different HDC group.
All Virgo galaxies are found at close to the same $\delta_g$, because the 5.7 Mpc smoothing scale of $\delta_g$ is larger than the 3.5 Mpc size of Virgo as defined for \atd\ galaxies.
The fractions of \atd\ galaxies in each $\nu_{10}$ bin does not add up to 1, because our $\nu_{10}$ is calculated using a parent sample of $M_K < -23.0$ (see \aref{sec:morenu} for details).
}
\label{fig:mhdeltanu}
\end{figure*}

The group catalogues of \citet{crooketal2007} also include dynamical estimates of dark matter halo masses for groups with at least 3 members. 
Two measurements of each halo are listed, one based on the standard virial estimator and the other based on the projected mass estimator \citep{heisleretal1985}. 
We use the latter since the former is sensitive to close pairs and can be noisy for groups not uniformly sampled spatially \citep[e.g.][]{bahcalltremaine1981}.
Uncertainties on the projected mass estimator can be up to 0.5 dex for groups with only a few members, though they become smaller for groups with many members \citep{heisleretal1985}.

For the MASSIVE galaxies, 89 are in groups with 3 or more members and have available $\mh$ from the HDC catalogue.
For the 258 \atd\ galaxies not observed with MASSIVE, 158 are in groups with $M_{\rm halo}$ measurements in the same catalogue.
Additional halo mass measurements based on more detailed analyses are available for the three well-studied clusters of Virgo, Coma, and Perseus. 
For Virgo, we use $\mh = 5.5 \times 10^{14} M_\odot$ \citep[same as in][]{durrelletal2014}, which is a combination of the Virgo A and B subcluster masses \citep{ferrareseetal2012} and the M86 subcluster mass \citep{schindleretal1999}.
For Coma, we use $\mh = 1.8 \times 10^{15} M_\odot$, an average between $2.7 \times 10^{15} M_\odot$ \citep{kuboetal2007} from weak gravitational lensing measurements and $9.2 \times 10^{14} M_\odot$ (\citealt{falcoetal2014}, see also \citealt{rinesetal2003}) from galaxy dynamics. For Perseus, we use $\mh = 6.7 \times 10^{14} M_\odot$ from spatially-resolved Suzaku X-ray observations \citep{simionescuetal2011}.

The left panel of \mautoref{fig:mhdeltanu} compares the distribution of $\mh$ in the two surveys.
Nearly 40\% of MASSIVE galaxies are in haloes above $10^{14} M_\odot$, whereas only $\sim 5$\% of \atd\ galaxies {\it outside} of the Virgo region are in such massive haloes.  
The leftmost bin shows "isolated" galaxies with fewer than 3 group members in the HDC catalogue and hence with no available $\mh$ measurements; a higher fraction of \atd\ galaxies belong to this category than MASSIVE galaxies ($\sim 40$\% versus 23\%; see also \mautoref{sec:bgg} and \mautoref{fig:bgg}).

\subsection{Large-scale density}
\label{sec:delta}

The group membership and group halo mass that we have investigated thus far provide information about galaxy environment on scales of a few hundred kpc to $\sim 1$ Mpc.
Another useful measure of galaxy environment is the large-scale density field surrounding a galaxy on the scale of several Mpc.
To this end, we use the density field of \citet{carricketal2015} constructed from the 2M++ redshift catalogue of \citet{lavauxhudson2011}.
The 2M++ catalogue contains 69,160 galaxy redshifts from 2MRS, the Sloan Digital Sky Survey Data Release 7 \citep[SDSS-DR7;][]{abazajianetal2009}, and the 6dF galaxy redshift survey Data Release 3 \citep[6dFGRS-DR3;][]{jonesetal2009}.
It covers nearly the full sky and reaches a depth of $K = 12.5$ mag, deeper than $K = 11.75$ mag for the 44,599 galaxies in 2MRS alone.

\citet{carricketal2015} presents a luminosity-weighted galaxy density contrast, $\delta_g \equiv (\rho_g - \overline{\rho_g})/\rho_g$, smoothed with a 5.7 Mpc Gaussian kernel.
This density field is computed with weights assigned to each galaxy's luminosity to account for the magnitude limit of the survey and incompleteness.
It is also rescaled to account for the impact of luminosity dependent galaxy-matter bias on the density field calculated at different redshifts.
The result is a smoothed density field complete out to a distance of 178 Mpc (and partial coverage to a further distance of 286 Mpc).
Our survey (out to 108 Mpc) is well within this radius.

The grid spacing of the published density field is approximately 2.2 Mpc.
We compute the density at the location of each galaxy using a simple trilinear interpolation, which results in interpolation errors of approximately 0.1 dex.
For galaxy distance, we use the LDC group-corrected distance where available, and redshift distances from \mbox{\citet{huchraetal2012}} otherwise (assuming $h=0.73$ as in \citealt{crooketal2007}).
We use these distances here instead of the distances from \mbox{\citet{maetal2014}} because they are more comparable to the reconstruction procedure in \citet{carricketal2015}, but uncertainties in distance estimates will result in uncertainties in the density.

The middle panel of \mautoref{fig:mhdeltanu} shows the distributions of $\delta_g$ for the entire MASSIVE versus \atd\ sample.
The values of $\delta_g$ for the 75 MASSIVE galaxies with stellar kinematics are listed in \mautoref{table:main}.
While all MASSIVE galaxies are in regions above or near the cosmic mean density, about 20\% of \atd\ galaxies are in underdense or mean-density regions.  The Coma cluster and Perseus cluster are the two highest-density regions sampled by the MASSIVE survey, both
with $\delta_g \approx 12$.  In comparison, the Virgo cluster is the highest-density region sampled by the \atd survey with
$\delta_g \approx 8$ (dotted line in \mautoref{fig:mhdeltanu}).  
Because $\delta_g$ is smoothed over a scale larger than the size of a galaxy cluster, all galaxies in the same cluster
have the same $\delta_g$.  

\mautoref{fig:bigmap} is a sky map of $\delta_g$ contours over the MASSIVE volume.  The MASSIVE galaxies (black circles)
are located in regions with $\delta_g \ga 0$ (yellow and light green), whereas many \atd\ galaxies (grey squares) are in
the Virgo cluster or lower density regions (dark green).  As expected, the parent sample of $\sim 15,000$ early-type galaxies 
with $M_K < -23.0$ mag from 2MRS (white dots) traces the $\delta_g$ contours quite well.  

\subsection{Local density}
\label{sec:nu}

Finally, we calculate a local galaxy density by finding the distance to the $N$th nearest neighbour of a galaxy and estimating the luminosity (or mass) enclosed in this region.
Several versions of local densities were tabulated in \citet{cappellarietal2011b} for the \atd\ sample.
We will focus on $\nu_{10}$, the luminosity density of galaxies in a sphere enclosing the $10^{\rm th}$ nearest neighbour (where the galaxy itself is counted as the 0$^{\rm th}$ neighbour).

The galaxy sample used to estimate $\nu_{10}$ in \citet{cappellarietal2011b} included all 2MRS galaxies (including spirals) with $M_K < -21.5$ mag in the \atd\ volume.
This cut matches the completeness limit $K = 11.75$ mag of 2MRS, which corresponds to $M_K \approx -21.5$ mag at the edge of the \atd\ volume (42 Mpc).   
At the edge of the much larger volume probed by MASSIVE (108 Mpc), however, 2MRS is complete only to $M_K \approx -23.4$ mag.
Using the same parent sample as in \atd\ to calculate $\nu_{10}$ for MASSIVE galaxies would thus suffer substantially from incompleteness.
Instead, We choose a magnitude cut of $M_K=-23.0$ mag for defining the parent 2MRS sample and compute $\nu_{10}$ from this sample of approximately $10^4$ galaxies for both MASSIVE and \atd\ galaxies, for a fair comparison between the two surveys.
For simplicity and uniformity, we use HDC group distances (where available; LDC group distances otherwise, and redshift distances as a last resort) for all galaxies in this calculation.
This includes the survey galaxies, even if they have more accurate distances tabulated in \citet{maetal2014} or \citet{cappellarietal2011a}.
Using group distances results in typical uncertainties of about 0.5 dex on $\nu_{10}$, due to individual galaxy distances being flattened to the group distace.
Rarely, $\nu_{10}$ may be inflated by more than this because the distance flattening also re-orders the closest neighbors, but this impacts only a handful of galaxies.
Additional details are discussed in \aref{sec:morenu}.

The distribution of the resulting $\nu_{10}$ for each survey is shown in the right panel of \mautoref{fig:mhdeltanu}.
The values of $\nu_{10}$ for the MASSIVE priority sample are listed in \mautoref{table:main}.  
We express $\nu_{10}$ in units of the mean $K$-band luminosity density $\overline{\nu} \sim 2.8 \times 10^8\; {\rm L}_\odot \; {\rm Mpc}^{-3}$ for magnitude ranges of $-21 > M_K > -25$ mag from Table~2 of \citet{lavauxhudson2011}.
As their Table~2 shows, enlarging the range to $-17 > M_K > -25$ mag would raise $\overline{\nu}$ by only 5\%, so the accuracy of the magnitude range is not a concern.

\mautoref{fig:mhdeltanu} and \autoref{table:main} show that MASSIVE galaxies span about five orders of magnitude in $\nu_{10}$, reaching
$\nu_{10}$ above $10^4$ for galaxies at the centres of the Coma cluster, Perseus cluster, Virgo cluster, Abell 194, Abell 262, Abell 347, and Abell 1367.

An alternate measure of local density is $\Sigma_3$, defined as the number density of galaxies in a cylinder of 600 \kms\ in height, enclosing the 3 nearest neighbours, centred on the galaxy.
This avoids the requirement of good redshift-independent distances by replacing it with a flat cutoff requiring neighbours to have heliocentric velocities within 300 \kms\ of the central galaxy.
It is more sensitive to issues of survey completeness, and the overall results using $\Sigma_3$ are not much different than those using $\nu_{10}$, so we do not discuss it in the body of the paper.
For completeness, we do present results in \aref{sec:atd} for the original $\Sigma_3$ and $\nu_{10}$ densities calculated in \citet{cappellarietal2011a}.

\subsection{Relationships among different measures of environment}
\label{sec:env-env}

\begin{figure}
\begin{center}
\includegraphics[page=4]{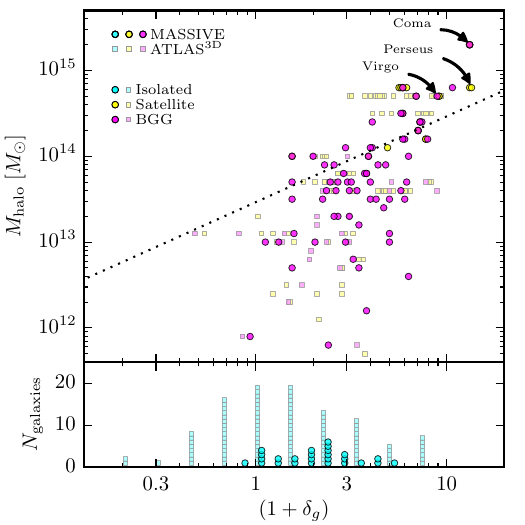}
\end{center}
\caption{
(Top) Halo mass  $\mh$ versus large-scale density  field $1+\delta_g$ for MASSIVE (circles) and \atd\ (squares) galaxies, where
BGGs (magenta) and satellites (yellow) are colour-coded separately.
Most satellite galaxies are hidden behind their respective BGGs since they have identical or very similar $\delta_g$.
The dotted line shows the extreme case when the halo mass dominates the total mass within the volume $V$ of a sphere with radius of the smoothing distance (5.7 Mpc) used to measure $\delta_g$, i.e., $\mh = (1+\delta)\,\overline{\rho}\,V$. Along this line, $\delta_g$ is simply measuring $\mh$; away from this line, $\delta_g$ and $\mh$ offer independent measures of a galaxy's environment.
(Bottom) Galaxies with fewer than 3 group members in the HDC catalogue are classified as isolated and have no estimated halo mass; the distribution of their $\delta_g$ are shown (cyan).
}
\label{fig:mh-delta}
\end{figure}

\begin{figure}
\begin{center}
\includegraphics[page=5]{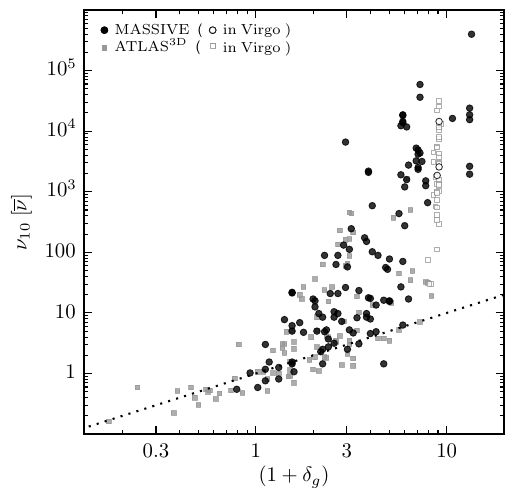}
\end{center}
\caption{Local density $\nu_{10}$ versus large-scale density $1+\delta_g$ for MASSIVE (black circles) and \atd\ (grey squares) galaxies.
  At low densities, $\nu_{10}$ and $(1+\delta_g)$ follow the dotted one-to-one line almost exactly, as expected when the $10^{\rm th}$ neighbour is at a distance comparable to the smoothing scale for $\delta_g$.
At high densities, e.g., within the Virgo cluster (open symbols), $\nu_{10}$ measures the local galaxy density and spans a much larger range than $\delta_g$.
}
\label{fig:nu-delta}
\end{figure}

Here we examine how the environmental measures discussed above -- group membership, $\mh$, $\delta_g$, and $\nu_{10}$ -- correspond to one another.
The distributions in $\delta_g$ for isolated galaxies in the two surveys are shown at the bottom of \mautoref{fig:mh-delta}.
The top panel of \mautoref{fig:mh-delta} shows $\mh$ vs $\delta_g$ for BGGs (magenta) and satellites (yellow), where higher-mass haloes generally reside in higher-density regions.  
This is expected since clusters of mass above $\sim 10^{14} M_\odot$ dominate the overdensity within 5.7 Mpc, the smoothing scale of $\delta_g$.
If a galaxy halo with $\mh$ were the only mass within the smoothing volume $V$
(of radius 5.7 Mpc), the overdensity would be
$\mh = (1+\delta)\overline{\rho}\,V$, as indicated by the dotted line in 
\mautoref{fig:mh-delta}. The galaxies with $\mh \ga 10^{14} M_\odot$ 
indeed lie near this line.  Many lower-mass halos, on the other hand,
lie to the right of this line since $\delta_g$ measures the large-scale density
field rather than the halo mass itself in this regime;
$\delta_g$ therefore spans a wide range of values, from near the cosmic mean to regions with $\delta_g$ near 10.

Satellite galaxies are sometimes visible as ``tails'' to the left of the BGG galaxy in the same halo in \mautoref{fig:mh-delta}, when the halo is large enough for the outskirts to show noticeably lower density on the smoothing scale of $\delta_g$.
Many more satellite galaxies, however, are hidden behind their BGG galaxies on this figure; in particular, all galaxies defined as Virgo galaxies by \atd\ are within one symbol-width of the Virgo BGG.
The other satellites defined as part of the same halo by the HDC catalogue were not designated as Virgo galaxies in \atd.

\mautoref{fig:nu-delta} shows the relationship between local density $\nu_{10}$ and large-scale density $\delta_g$.
At low densities, where both the distance to the $10^{\rm th}$ neighbour and the smoothing scale of $\delta_g$ are at scales beyond the size of the host halo, $\nu_{10}$ follows $\delta_g$ almost exactly.
In dense regions, however, $\nu_{10}$ and $\delta_g$ can deviate significantly,
where $\nu_{10}$ is determined by the innermost 10 galaxies well within a single host halo, while $\delta_g$ continues to measure the overdensity on several Mpc scale surrounding the galaxy.
The Virgo galaxies (open symbols) illustrate this difference, showing a large spike in $\nu_{10}$ relative to $\delta_g$.
Perseus shows the highest $\nu_{10}$, and other groups/clusters (including Coma) in dense environments are similarly far above the smoothed density $\delta_g$.


\section{Galaxy spin versus stellar mass}
\label{sec:lam}

In this section we present measurements of rotation in MASSIVE galaxies and investigate the dependence of galaxy spin on $M_*$ and $M_K$ in the \atd\ and MASSIVE surveys.
Here, and throughout the rest of the paper, we consider the 75 galaxies of the priority sample ($M_K < -25.5$ mag, $M_* \ga 10^{11.7} \; M_\odot$) of the MASSIVE survey, for which we have completed the Mitchell IFS observations.
This sample nearly doubles the 41 galaxies brighter than $M_K = -25.7$ mag reported in \citet{vealeetal2017}.

\subsection{Galaxy spin $\lambda_e$}
\label{sec:lam-def}

\begin{figure}
\begin{center}
\includegraphics[page=1]{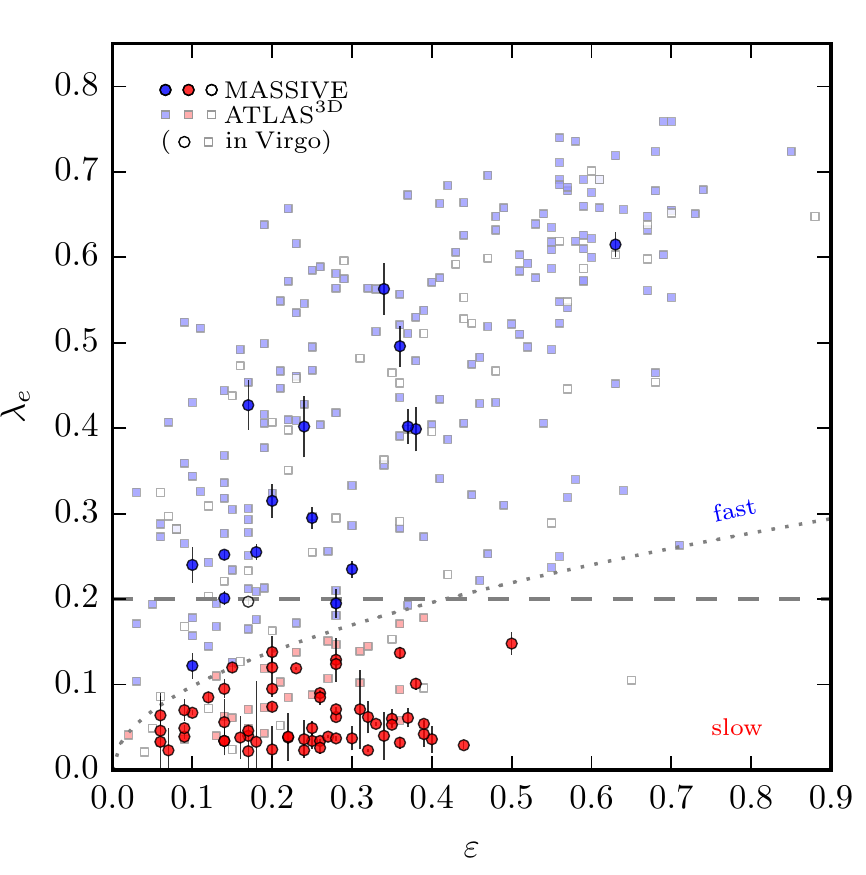}
\end{center}
\caption{Spin parameter $\lambda_e$ versus ellipticity $\epsilon$ for MASSIVE (circles) and \atd\ (squares) galaxies. Open symbols indicate Virgo galaxies.
Slow (red) and fast (blue) rotators can be separated by either a flat cutoff (dashed line; \citealt{lauer2012}), or one that takes into account ellipticity \citep[dotted line;][]{emsellemetal2011}.
We use the latter cutoff in this analysis, but have tested that a flat cutoff does not qualitatively change the results.
}
\label{fig:lam-eps}
\end{figure}

As discussed in \mautoref{sec:samples}, our kinematic data provide measurements of the stellar velocity ($V$), dispersion ($\sigma$), and higher velocity moments ($h_N$) for each spatial bin folded across the major axis.
We use the dimensionless parameter $\lambda$ to quantify the relative importance of rotation in a galaxy:
\begin{equation}
  \label{eq:lambda}
  \lambda(<R) \equiv \frac{\langle R |V| \rangle}{\langle R \sqrt{V^2 + \sigma^2} \rangle} \,, \quad \lambda_e \equiv \lambda(<R_e)\,,
\end{equation}
where the angle brackets represent luminosity-weighted averages over all bins within $R$ \citep{emsellemetal2007}.  
The luminosity-weighting and cumulative nature of $\lambda(<R)$ prevents it from varying rapidly past $R_e$, unlike the local $\lambda(R)$ sometimes used to investigate radial gradients in rotation structure.
The fact that our $\lambda(<R)$ is largely flat by $R_e$ is important to minimize any bias in $\lambda_e$ due to possibly underestimated $R_e$ measurements \citep{vealeetal2017}.

The values of $\lambda_e$ for the 75 MASSIVE galaxies are plotted against ellipticity in \mautoref{fig:lam-eps} (circles) and are listed in \mautoref{table:main}.
The \atd\ sample is also shown (squares) for comparison.
We note that our measurements of $\lambda_e$ use circular bins \citep{vealeetal2017}, whereas \atd\ uses elliptical apertures.
Since most MASSIVE galaxies are quite round ($\varepsilon < 0.4$), the difference between circular and elliptical apertures is likely to be insignificant.
\citet{emsellemetal2011} classified galaxies as fast or slow rotators according to $\lambda_e = 0.31 \sqrt{\varepsilon}$, where all galaxies with $\lambda_e$ below this cutoff are defined as slow rotators.
A flat cutoff at $\lambda_e = 0.2$ had also been suggested \citep{lauer2012}, and could be appropriate for our galaxies for several reasons as discussed in \citet{vealeetal2017}.
We have tested that the conclusions of this paper are not changed if we use this flat cutoff, which by extension ensures that our conclusions are not impacted by uncertain measurements of $\varepsilon$.
For simplicity, we use only the \atd\ cutoff hereafter.

We note that the IFS observations of many \atd\ galaxies do not reach
$R_e$, so their tabulated $\lambda_e$ is calculated within a smaller radius than  in the MASSIVE survey. 
About half of \atd\ galaxies have data extending to between $0.5 R_e$ and $R_e$, and about 8\% have observations extending to less than $0.5 R_e$ \citep{emsellemetal2011}.
Because many galaxies show a rising $\lambda(<R)$ profile within $R_e$, this may result in under-estimated $\lambda_e$ for \atd\ galaxies, and perhaps in fast rotators being misclassified as slow rotators.
For a galaxy with $\lambda(<R)$ measured to $0.5 R_e$, the appropriate slow/fast cutoff would be reduced by a factor of $\sim 1.2$ \citep{emsellemetal2011}.
Based on this rough scaling, only 4 \atd\ galaxies may be misclassified as slow rotators.
One of these is NGC 4472 (discussed in detail in \citealt{vealeetal2017}), which has $\lambda_e = 0.077$ according to \citet{emsellemetal2011}, measured with data going to only $0.26 R_e$, and has $\lambda_e = 0.2$ in our sample.
Of the remaining overlap galaxies between MASSIVE and \atd, only NGC 5322 and NGC 5557 (below the $M_K < -25.5$ cut of this paper) have been observed, and they both agree to within about $\Delta \lambda_e \sim 0.02$ with the results of \atd.

Another source of uncertainty is the measured PA of the galaxies; because we fold our data over the photometric PA, any misalignment between the rotation axis and photometric minor axis would wash out the rotation.
NGC 1129 and NGC 4874 were misaligned in this way, so we manually adjusted the folding PA \citep[see][]{vealeetal2017}.
Based on those manual adjustments we estimate minor misalignments ($\sim 10$~degrees) to reduce $\lambda_e$ by $\sim 0.01$, although the impact may be greater for faster rotators.
Formal errors on $\lambda_e$, calculated from the uncertainties on $V$ and $\sigma$ in each bin according to the formulae in \citet{houghtonetal2013}, are shown in \autoref{fig:lam-eps} for MASSIVE galaxies.
They are generally smaller than the potential systematic uncertainties discussed here, with over half of MASSIVE galaxies having formal errors less than $0.01$.
Finally, the $V$ and $\sigma$ used to calculate $\lambda_e$ may differ by up to 10\% between Gaussian-only fits (as used by \atd) and Gauss-Hermite fits including higher moments (as we use) \citep{vandermarelfranx1993}.

\subsection{$\lambda_e$ versus $M_*$}
\label{sec:lam-mk}

\begin{figure}
\begin{center}
\includegraphics[page=2]{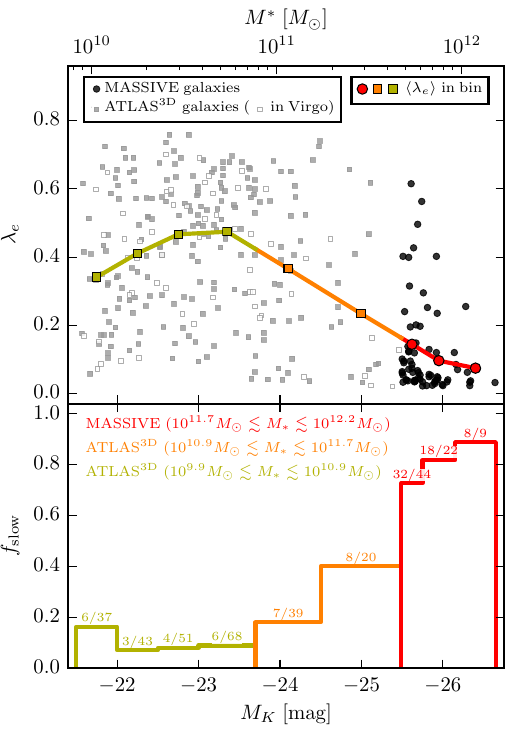}
\end{center}
\caption{Spin parameter $\lambda_e$ (top) and slow rotator fraction $f_{\rm slow}$ (bottom) versus $M_*$ (and $M_K$) for MASSIVE (black circles) and \atd\ galaxies (grey squares). The top panel shows that the MASSIVE and \atd\ galaxies span a similar wide range of $\lambda_e$ at $M_* \la 10^{12} M_\odot$, but the mean $\lambda_e$ (colour symbols) decreases sharply at high $M_*$.
Open symbols indicate Virgo galaxies.
We divide galaxies from the two surveys into three broad stellar mass bins indicated by red, orange and olive
for analysis in \mautoref{sec:lam-env}.
}
\label{fig:lam-mk}
\end{figure}

\begin{figure*}
\begin{center}
\includegraphics[page=1]{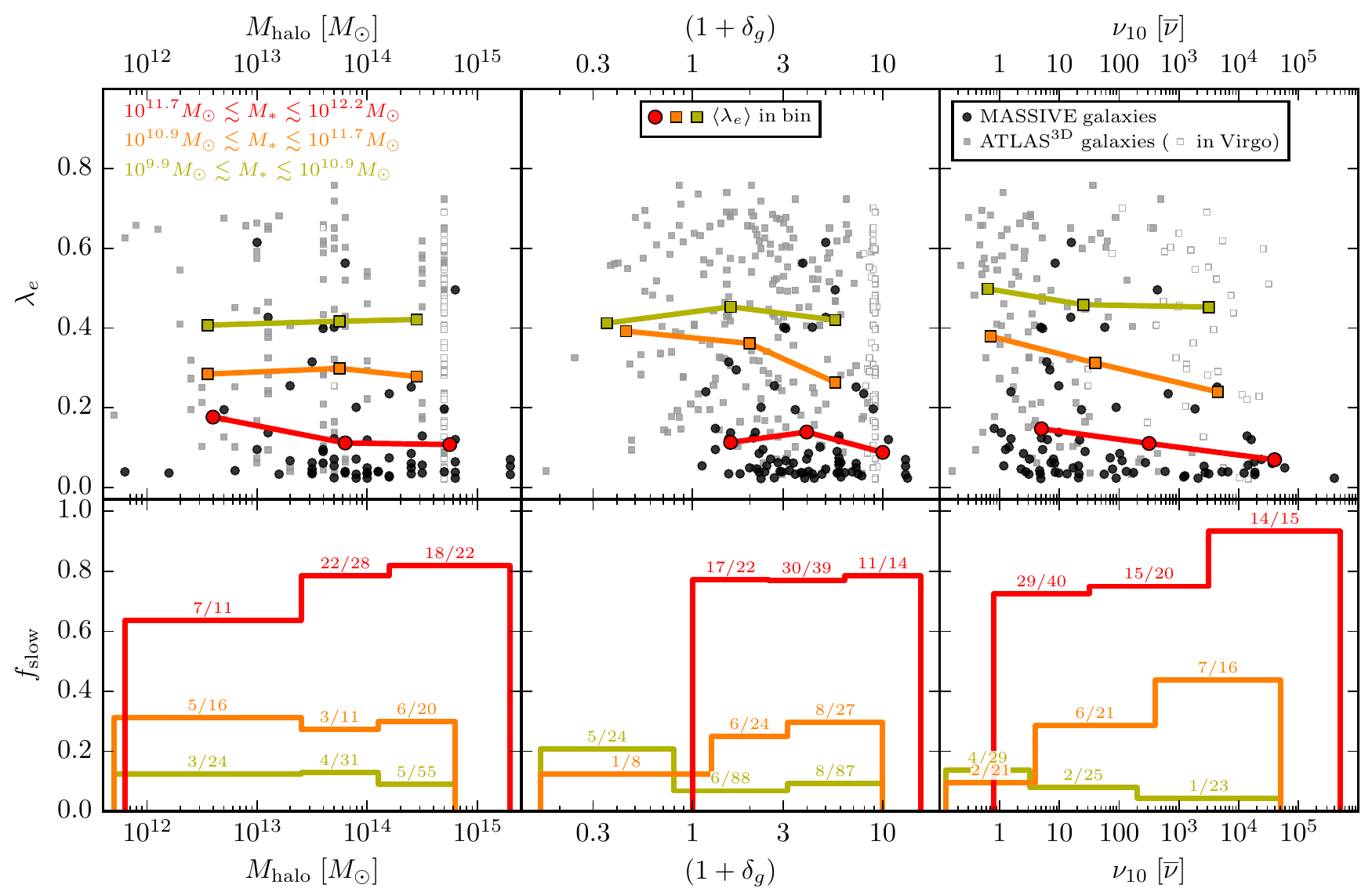}
\end{center}
\caption{
Spin parameter $\lambda_e$ (top) and slow rotator fraction $f_{\rm slow}$ (bottom) as a function of $\mh$ (left), $1+\delta_g$ (middle), and $\nu_{10}$ (right) for MASSIVE (black circles) and \atd\ (grey squares) galaxies. 
Colour symbols and lines show the average $\lambda_e$ (top) and $f_{\rm slow}$ (bottom) for three $M_*$ bins.
Overall, the average $\lambda_e$ and $f_{\rm slow}$ both depend strongly on $M_*$, as shown in \autoref{fig:lam-mk}.
Within each $M_*$ bin, however, 
both quantities vary little with $\mh$, $\delta_g$ or $\nu_{10}$,.
 The most noticeable trend with environment is 
a decrease in the average $\lambda_e$ (and an increase in $f_{\rm slow}$) with increasing $\nu_{10}$ for the middle (orange) mass bin in the right panel.
}
\label{fig:lam-real}
\end{figure*}

We find a strong anti-correlation between $\lambda_e$ and $M_*$, similar to our earlier finding from the smaller sample of 41 MASSIVE galaxies in \citep{vealeetal2017}.
The top panel of \mautoref{fig:lam-mk} shows $\lambda_e$ versus $M_*$ (and $M_K$) for each galaxy in MASSIVE and ATLAS$^{\rm 3D}$.
Even though the range of $\lambda_e$ at a given $M_*$ is similar for almost the entire range of  $M_*$, the average $\lambda_e$ over 9 $M_*$ bins (colour symbols) drops from $\sim 0.4$ for galaxies below $M_* \sim 10^{10.5} M_\odot$, to below 0.1 for galaxies above $M_* \sim 10^{12} M_\odot$.   The bottom panel of \mautoref{fig:lam-mk} plots the corresponding steep rise in the fraction of slow rotators $f_{\rm slow}$, from $\sim 10$\% to $\sim 90$\% with increasing $M_*$.
The average behaviour of $\lambda_e$ versus $M_*$ in the top panel does not substantially change if we normalize $\lambda_e$ by the slow/fast cutoff ($0.31 \sqrt{\varepsilon}$).
Some individual round galaxies have very high normalized $\lambda_e$, but the average behaviour is qualitatively very similar.

There is a slight decrease in average $\lambda_e$ and slight increase in $f_{\rm slow}$ at the low mass end of the \atd\ sample.
This is not due to any incompleteness of the sample \citep{cappellarietal2011a}, and also occurs if we use a flat cutoff to define the slow rotators, so it is not due to changes in $\epsilon$ influencing the classification.
Perhaps coincidentally, the peak in the ETG mass function at $M_K \sim -22.5$ mag is near the peak of $\langle \lambda_e \rangle$ and the minimum of $f_{\rm slow}$.
It is also near the inferred peak in star formation efficiency at approximately $10^{10.5} M_\odot$ \citep[e.g.][]{behroozietal2013}.
Since we are focusing on the highest mass galaxies, we will not discuss these trends further.

\section{Galaxy spin versus environment}
\label{sec:lam-env}

In this section we examine the relation between galaxy spin and the various galaxy environmental measures defined in \mautoref{sec:env} for the combined MASSIVE and \atd\ sample. 
\mautoref{sec:lam-real} and \mautoref{sec:lam-split} present simple tests of the correlation between slow rotator fraction $f_{\rm slow}$ and $\mh$, $\delta_g$, and $\nu_{10}$, and \mautoref{sec:lam-pred} presents a more detailed test of how the joint correlation of $M_*$ vs. $\lambda_e$, and $M_*$ vs. environment, impacts our results. 

\subsection{$\lambda_e$ versus halo mass and density}
\label{sec:lam-real}

\mautoref{fig:lam-real} shows the distribution of $\lambda_e$ (upper panels) and $f_{\rm slow}$ (lower panels) versus halo mass $\mh$ (left), large-scale density contrast $\delta_g$ (middle), and local density $\nu_{10}$ (right).
Measurements of $\lambda_e$ for individual MASSIVE (black circles) and \atd\ (grey squares) galaxies are shown in the  upper panels.
The colored lines show the average $\lambda_e$ and $f_{\rm slow}$ for three $M_*$ bins: $M_* > 10^{11.7} M_\odot$ (red) contains all MASSIVE galaxies in this study;
and the other two bins contain \atd\ galaxies with $10^{10.9} M_\odot < M_* < 10^{11.7} M_\odot$ (orange) and $10^{9.9} M_\odot < M_* < 10^{11.7} M_\odot$ (olive), respectively.

The average $\lambda_e$ in \mautoref{fig:lam-real} is seen to decrease with increasing $M_*$ bin, and
$f_{\rm slow}$ is seen to increase correspondingly, as discussed in the previous section.  
Within each $M_*$ bin, however, there is at most a weak correlation with environment.
The lack of correlation applies regardless of the exact quantity we used to measure spin (i.e., individual $\lambda_e$, $\langle \lambda_e \rangle$, $f_{\rm slow}$) or environment (i.e., $\mh$, $\delta_g$, $\nu_{10}$).
The MASSIVE galaxies occupy nearly the same range of $\lambda_e$, but have many more galaxies near $\lambda_e = 0$.
Those slow and non-rotating galaxies occupy the same range of environments as our overall sample, resulting in a low $\lambda_e$ and high $f_{\rm slow}$ in all environments for our high-mass galaxies.

Some subtle trends with environment may be seen, although none are obviously significant given the number statistics of our samples, with one or two galaxies being the margin of difference in many cases.
For MASSIVE galaxies, larger halo mass correlates with a slightly lower average $\lambda_e$ and slightly higher $f_{\rm slow}$ (red lines in the left panels of \mautoref{fig:lam-real}). 
Similar correlations also apply to $\nu_{10}$ for MASSIVE galaxies (red lines in the right panels of \mautoref{fig:lam-real}), and to $\delta_g$ and $\nu_{10}$ for the more massive half of \atd\ galaxies (orange lines in
the middle and right panels of \mautoref{fig:lam-real}). 

\begin{table}
\caption{$p$-values from KS test on slow and fast rotators.
Small $p$-values indicate slow and fast rotators are likely drawn from
different distributions in the given quantity.
For the three environment measures we run the KS test separately on
the two broad $M_K$ bins of ATLAS$^{\rm 3D}$; $M_K$ for each bin is
given in magnitudes below.}
\label{table:kstest}
\begin{center}
\begin{tabular}{lcccc}
 & $M_K$ & $M_{\rm halo}$ & $\delta_g$ & $\nu_{10}$ \\
\hline
${\rm MASSIVE}$$\; (< -25.5)$ & 0.314 & $0.305$ & $0.710$ & $0.414$ \\
${\rm ATLAS}^{\rm 3D}$$\; (-23.7 \; {\rm to} \; -25.5)$ & \multirow{2}{*}{ 0.007 } & $0.835$ & $0.156$ & $0.112$ \\
${\rm ATLAS}^{\rm 3D}$$\; (> -23.7)$ &  & $0.765$ & $0.572$ & $0.316$ \\
\hline
combined & $10^{-8}$ & - & - & -
\end{tabular}
\end{center}
\end{table}

\begin{figure*}
\begin{center}
\includegraphics[page=1]{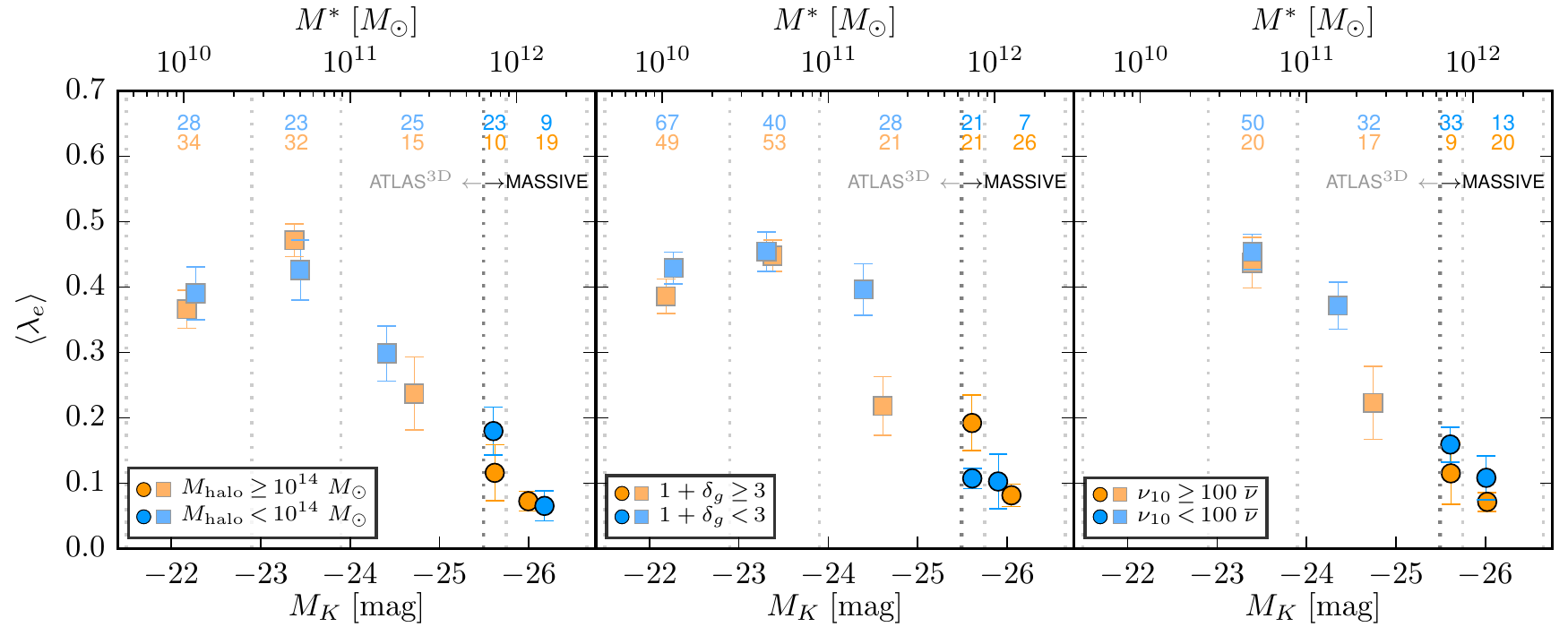}
\end{center}
\caption{Average $\lambda_e$ in bins of $M_*$ (demarcated by dotted lines) for MASSIVE (circles) and \atd\ (squares) galaxies.
Within each $M_*$ bin, the average of $\lambda_e$ is taken over two subsamples, one for galaxies in denser environments (orange) and one for less dense environments (blue); the corresponding number of galaxies is listed at the top of each bin.
 From left to right, the panels show splits by $\mh$, $\delta_g$, and $\nu_{10}$.
 (We do not calculate $\nu_{10}$ for galaxies with $M_K > -23.0$ mag; see \aref{sec:morenu} for details.)
Each $\langle\lambda_e\rangle$ point is placed at the average $M_*$ for the galaxies in that environmental bin.
While the orange point is significantly lower than the blue point in the highest mass \atd\ bin ($10^{11} M_\odot \la M_* \la 10^{11.7}M_\odot$), it is also at a higher average $M_*$ than the blue point.
Altogether, the decrease in the average $\lambda_e$ with $M_*$ continues smoothly from the \atd\ sample to the MASSIVE sample.  
There is no evidence that galaxies at the same $M_*$ but with different environments have different rotation properties.
}
\label{fig:lam-split}
\end{figure*}

To quantify the significance of these trends, we run two-sample Kolmogorov-Smirnov (KS) tests to compare the distribution of slow rotators to the distribution of fast rotators in $\mh$, $\delta_g$, and $\nu_{10}$.\footnote{Because the KS test is not suitable for discrete parameters, the large number of \atd\ galaxies at the same $\mh$ in Virgo (and to a lesser degree the duplicated $\mh$ values of other haloes for both surveys) causes a problem.
To solve this, we add a small random variable between $\pm 0.1$ to $\log_{10} \mh$ before computing the KS test, and run 1000 trials of this procedure to find the average $p$-value.}
The resulting $p$-values for MASSIVE and \atd\ are listed in \mautoref{table:kstest}.
Many values are greater than 0.5, indicating that it is more likely than not that the slow and fast rotators are drawn from identical distributions in environment.
The smallest $p$-values for environment are $\sim 0.1$ for the distributions in $\delta_g$ and $\nu_{10}$ of the more massive half of the \atd\ galaxies.
These align with the qualitative trends we noted above, but are still not considered significant.
In comparison, a KS-test for the distribution of slow and fast rotators with $M_K$ gives $p = 0.007$ for the \atd\ sample, $p = 0.31$ for the MASSIVE sample, and approximately $p \sim 10^{-8}$ for a combined sample.\footnote{For the combined sample, we copy each \atd\ galaxy 10 times before finding the KS statistic to account for the fact that the MASSIVE volume is approximately 10 times larger.
  This gives a reasonable overall distribution in $M_K$, with no kink in the cumulative distribution function to inflate differences between the slow and fast rotators.
To convert the KS statistic into a $p$-value, we use the original sample sizes, so the $p$-value is not artificially small due to artificially large $N$.}

\subsection{$\lambda_e$ versus $M_*$, for two environmental bins}
\label{sec:lam-split}

In the previous subsection we examined $\lambda_e$ as a function of environment for three $M_*$ bins.
Here we investigate $\lambda_e$ as a function of $M_*$ for a low-density versus a high-density environmental bin.

\mautoref{fig:lam-split} shows the average $\lambda_e$ versus $M_*$, split into two samples (orange versus blue) by each of our three environment measures: $\mh$ (left), $\delta_g$ (middle), and $\nu_{10}$ (right).
The trend of $\lambda_e$ with $M_*$ for each environmental group in \mautoref{fig:lam-split} follows closely the trend of the whole sample in \mautoref{fig:lam-mk}.
Within a $M_*$ bin, however, there is again no statistically significant difference in the average $\lambda_e$ for the higher versus lower density sample for any of the three environmental variables.
In the central $M_*$ bin, which spans a factor of $\sim 5$ in $M_*$ ($10^{11} M_\odot < M_* < 10^{11.7} M_\odot$), the galaxies in the higher-density sample (orange) have a slightly higher average $M_*$ and a lower average $\lambda_e$ than the lower-density sample (blue). 
These points continue smoothly into the highest mass bins populated by the MASSIVE galaxies.

Like \autoref{fig:lam-real}, \mautoref{fig:lam-split} again illustrates that for local ETGs {\it of similar $M_*$}, the galaxy spin does not correlate noticeably with galaxy environment.

\subsection{$\lambda_e$, $M_*$, and environment}
\label{sec:lam-pred}

\begin{figure*}
\begin{center}
\includegraphics[page=4]{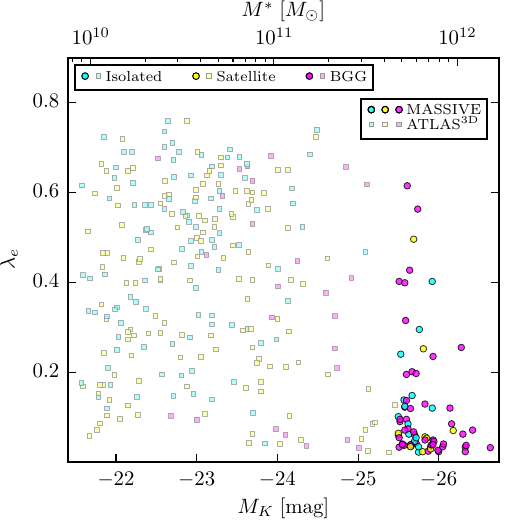}
\includegraphics[page=3]{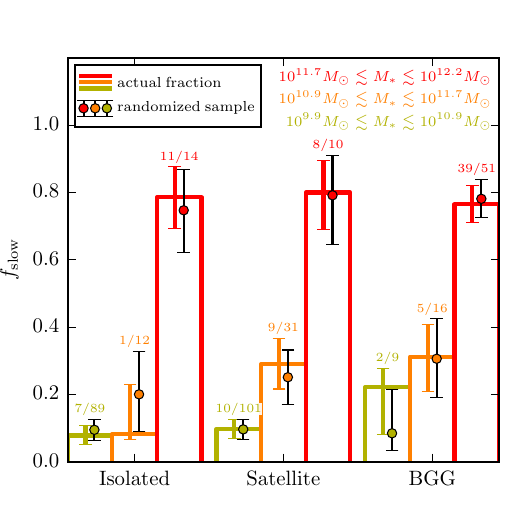}
\end{center}
\caption{$\lambda_e$ versus $M_*$ (left) and $f_{\rm slow}$ (right) divided by group membership status for MASSIVE and \atd\ galaxies.
  BGG galaxies (magenta) tend to have higher $M_*$ than isolated (cyan) and satellite (yellow) galaxies.
Combined with the increasing $f_{\rm slow}$ at higher mass, this results in a higher $f_{\rm slow}$ (right panel) among BGG galaxies in the more massive \atd\ mass bin (orange).
The trend holds, within errors, even when the slow/fast classification of each galaxy is randomized within each $M_*$ bin (symbols with black error bars; see text for details).
}
\label{fig:lam-bgg}
\end{figure*}

\begin{figure*}
\begin{center}
\includegraphics[page=2]{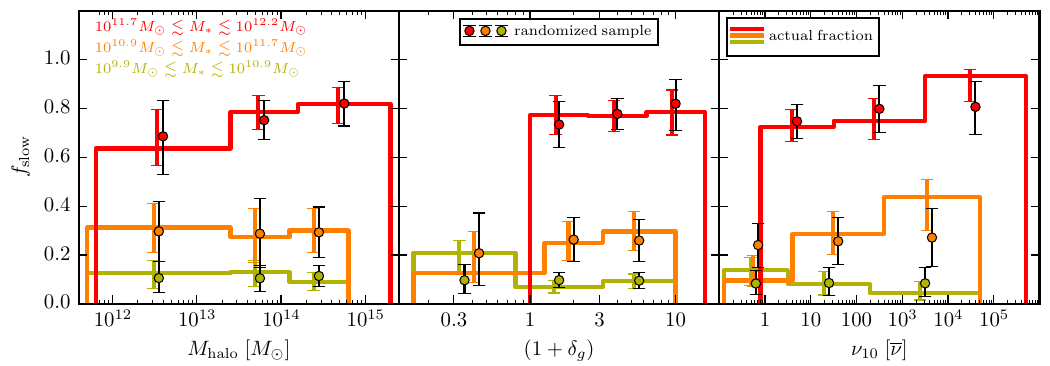}
\end{center}
\caption{Same as the bottom panels of \autoref{fig:lam-real} for $f_{\rm slow}$ versus $\mh$ (left), $\delta_g$ (middle), and $\nu_{10}$ (right).
In each panel, the randomized test sample (symbols with black error bars) matches very well the observed $f_{\rm slow}$ (histograms with colour error bars).
A possible exception is that the observed $f_{\rm slow}$ may have a steeper correlation with $\nu_{10}$ than the randomized sample, for the MASSIVE galaxies (red) and the more massive half of the \atd\ galaxies (orange).
}
\label{fig:lam-pred}
\end{figure*}

We perform an additional test to gain further insight into the lack of correlation between galaxy spin and environment at fixed stellar mass. 
We create a randomized sample in which the galaxies' $M_*$ and environment are preserved but the spins are shuffled.  
Specifically, we randomly assign each galaxy to be a fast or slow rotator, using the probability for being a slow rotator determined by the galaxy's $M_*$ shown in \mautoref{fig:lam-mk} (bottom panel).
For example, a galaxy with $M_* = 10^{10.6} M_\odot$ has a 6/68 = 8.8\% chance of being a slow rotator,
and a galaxy with $M_* = 10^{12.1} M_\odot$ has a 8/9 = 89\% chance of being a slow rotator, independent of their environment. 
We then count the fraction of slow rotators in each environment for each survey, and repeat the procedure 1000 times to obtain an estimate of the fraction and error bar \footnote{The 68\% confidence interval based on the cumulative distribution function of trials.} that reflects expected errors due to small sample sizes. 
We estimate error bars on the observed $f_{\rm slow}$ with a Bayesian method described in \aref{sec:err}.

\mautoref{fig:lam-bgg} shows $f_{\rm slow}$ for this test sample (symbols with black error bars)
for three $M_*$ bins for the three types of group memberships (isolated, satellite and BGG). 
The fact that the randomized test sample reproduces the observed $f_{\rm slow}$ supports our finding of the lack of an independent correlation between galaxy spin and environment as quantified by group membership.
The much higher fraction of slow rotators in MASSIVE (red) than \atd galaxies (orange and olive) seen in \mautoref{fig:lam-bgg} arises from their higher $M_*$ and not galaxy environment.
Likewise, the increasing fraction of slow rotators with group membership category for \atd\ galaxies can be accounted for entirely (within errors) by the joint correlations between $M_*$ and spin, and between $M_*$ and environment.

\mautoref{fig:lam-pred} shows the same comparison as \mautoref{fig:lam-bgg} but for the other environmental measures.
The observed trends of $f_{\rm slow}$ with environment for the three $M_*$ bins is again reproduced in the test sample despite the randomization of galaxy spin.
One possible exception is the highest $\nu_{10}$ bin for the middle and high mass bins (orange and red in the right panel of \mautoref{fig:lam-pred}), where the test sample underpredicts the observed $f_{\rm slow}$ slightly.
For completeness, we also apply this procedure to the local densities $\nu_{10}$ and $\Sigma_3$ originally tabulated in \citet{cappellarietal2011b} (see \aref{sec:atd}) and obtain similar results.

\mautoref{fig:lam-bgg} and \mautoref{fig:lam-pred} show that the observed $f_{\rm slow}$ match closely the test samples in which galaxy spin and environment are randomized.  We therefore conclude that there is likely little or no correlation between spin and environment {\em for galaxies of the same mass}, and the apparent kinematic morphology-density is largely driven by stellar mass.

\section{Conclusions}
\label{sec:summary}

In this paper we have analysed the detailed environmental properties of the 116 galaxies in the MASSIVE survey and the 260 galaxies in the \atd\ survey.
The MASSIVE survey is designed to sample a much larger volume than \atd\ and to target exclusively galaxies above $M_* = 10^{11.5} M_\odot$; only 6 \atd\ galaxies are above this mass limit (\autoref{sec:samples}).
These two complementary IFS surveys together span $-21.5 \ga M_K \ga -26.6$ mag, or $8\times 10^9 \la M_* \la 2\times 10^{12} M_\odot$, and provide the most comprehensive study to date of individual early-type galaxies in the local universe. 

We examined different ways to quantify galaxy environment and presented results for group membership (BGG, satellite or isolated), halo mass, large-scale density $\delta_g$ measured over a few Mpc, and local density $\nu_{10}$ measured within the 10th nearest neighbour of each galaxy (\autoref{sec:env}). 
Despite their high stellar masses, MASSIVE galaxies reside in a diverse range of environments,
and not all of them are the brightest galaxies in massive haloes at high densities, as is commonly assumed for massive ETGs.
This key feature of massive galaxies was highlighted in \citet{maetal2014}; here we have added environmental measures $\delta_g$ and $\nu_{10}$ to the analysis.  
About 20\% of MASSIVE galaxies are ``isolated'', having fewer than three group members in the 2MASS HDC catalogue (\mautoref{fig:bgg}); about 30\% of MASSIVE galaxies are in regions of modest densities $\nu_{10} \la 10 \bar{\nu}$, or $\delta_g \la 2$ (\mautoref{fig:mhdeltanu}).
Compared to \atd\ galaxies, we found a higher fraction of MASSIVE galaxies to be BGGs ($\sim 60$\% versus 10\%) and to be located in more massive haloes and higher density regions (\mautoref{fig:bgg} and \mautoref{fig:mhdeltanu}).

We then presented measurements of galaxy spin $\lambda_e$
for the 75 galaxies of the MASSIVE ``priority sample'' with IFS data (\autoref{sec:lam}).  
We confirm the strong anti-correlation between
spin and stellar mass that we reported earlier for a smaller sample of MASSIVE galaxies \citep{vealeetal2017}:
the average $\lambda_e$ decreases from $\sim 0.4$ at $M_* \sim 10^{10} M_\odot$ to below 0.1 at $M_* \ga 10^{12} M_
\odot$, and the fraction of slow rotators increases from $\sim 10$\% to 90\% (\autoref{fig:lam-mk}).

The combined MASSIVE and \atd\ sample is sufficiently large for us to analyze for the first time the correlation between galaxy spin and environment -- the so-called kinematic morphology-density relation -- at {\it fixed} $M_*$.  
After controlling for $M_*$, we found almost no remaining correlations between galaxy environment and spin (\mautoref{fig:lam-real}- \mautoref{fig:lam-pred}).

In particular, we find a high fraction of slow rotators ($\sim 80$\%) in the MASSIVE sample in every environment, regardless of group membership, halo mass, or densities $\delta_g$ or $\nu_{10}$ (\mautoref{fig:lam-real}, \mautoref{fig:lam-bgg} and \mautoref{fig:lam-pred}).
Previous studies did not include high-mass galaxies at low environmental densities since they either sampled smaller volumes \citep{cappellarietal2011b} or investigated only individual clusters \citep{deugenioetal2013,houghtonetal2013,fogartyetal2014}.
Strong correlations between slow rotator fraction and local density have been reported for these samples, with possible implications for slow rotator formation pathways (see, e.g. \citealt{cappellari2016} for a review).
However, without the high-mass low-density galaxies like those in the MASSIVE survey, it was difficult to assess whether
the kinematic morphology-density relation applies at fixed mass, i.e., whether environmental overdensities play a role in the formation of slow rotators that is distinct from their role as a natural environment for massive ETGs in general.
The results presented in this paper show that a dense environment is not required to create massive slow rotators.

Our tests show that most observed correlations between galaxy environment and spin can be explained by the joint connections between $M_*$ and spin, and between $M_*$ and environment (\mautoref{fig:lam-split}, \mautoref{fig:lam-pred}).
This is consistent with a scenario in which mergers generally are responsible for both increasing the mass of a galaxy and decreasing the spin.
A possible exception is the local density $\nu_{10}$, which shows evidence that the highest densities host a slightly larger fraction of slow rotators, even after controlling for $M_*$ (\mautoref{fig:lam-pred}).
This residual dependence may indicate that certain types of assembly history -- perhaps those including more minor mergers and non-merger interactions as suggested by simulations (e.g., \citealt{moodyetal2014,choiyi2017}) -- are more likely to create slow rotators, even when controlling for the final mass of the galaxy, and that local density is a reasonable proxy for the type of assembly history.

The fast-slow kinematic transformation \citep[e.g. this work,][]{cappellari2013} can be compared to the spiral-elliptical morphological transformation \citep[e.g.][]{dressler1980}.
Galaxy kinematics and morphologies both transform with galaxy mass, so it is important to examine whether the transformation with environment applies at fixed mass.
The morphology-density relation nearly disappears for galaxy samples at fixed stellar mass (e.g., \citealt{bamfordetal2009}; \citealt{tascaetal2009}; \citealt{blantonmoustakas2009}; \citealt{grutzbauchetal2011}; \citealt{muzzinetal2012}; \citealt{alpaslanetal2015}; \citealt{saraccoetal2017}).
We have found the same to be true for the kinematic morphology-density relation, which, at fixed stellar mass, disappears completely for every environmental measure studied in this paper except perhaps $\nu_{10}$.  A recent pre-print finds similar results for galaxies in eight clusters with halo mass above $10^{14.2} M_\odot$ from the SAMI survey, with no significant remaining relationship between the slow rotator fraction and local overdensity after controlling for the strong correlation with mass \citep{broughetal2017}.

Increased statistics from ongoing and future surveys using IFS such as SAMI \citep{croometal2012}, CALIFA \citep{sanchezetal2012}, MaNGA \citep{bundyetal2015}, and HECTOR \citep{bryantetal2016} could provide more sensitive probes of the transition regime between fast and slow rotating ETGs.
The MASSIVE survey is designed to explore new parameter space unprobed by \atd; the two samples therefore have little overlap.
It is somewhat a coincidence that ETGs transition from being dominated by fast rotators to being dominated by slow rotators at $M_K \sim -25$ mag, the interface between the two surveys (\mautoref{fig:lam-mk}).
A volume-limited survey targeting more galaxies brighter than 
$M_K \sim -24$ mag would be useful for gaining further insight into the kinematic transformation along the mass sequence of present-day ETGs.


\section*{Acknowledgements}

We thank Mike Hudson for his assistance with the 2M++ catalogue.
The MASSIVE survey is supported in part by NSF AST-1411945,  NSF AST-1411642, HST-GO-14210, and HST-AR-1457.




\bibliographystyle{mnras}
\bibliography{massive_vii} 




\appendix

\section{Calculation of $\nu_{10}$}
\label{sec:morenu}

The local luminosity density $\nu_{10}$ was described briefly in \mautoref{sec:nu}.
Here we discuss some of the technical details of calculating $\nu_{10}$, which is defined as follows:
\begin{equation}
  \label{eq:nu}
  \nu_{10} = \frac{\Sigma_{i=0}^{10} 10^{-0.4(M_{i,K} - M_{\odot, K})}}{\frac{4}{3} \pi r_{10}^3}
\end{equation}
where the solar $K$-band luminosity is $M_{\odot,K} = 3.29$ mag \citep{blantonroweis2007}.
Index $i=0$ to 10 refers to the galaxy itself ($i=0$) and its ten nearest neighbours, so $r_{10}$ is the distance to the 10$^{\rm th}$ neighbour and defines a sphere containing the 10 neighbours.

\begin{figure}
\begin{center}
\includegraphics[page=1]{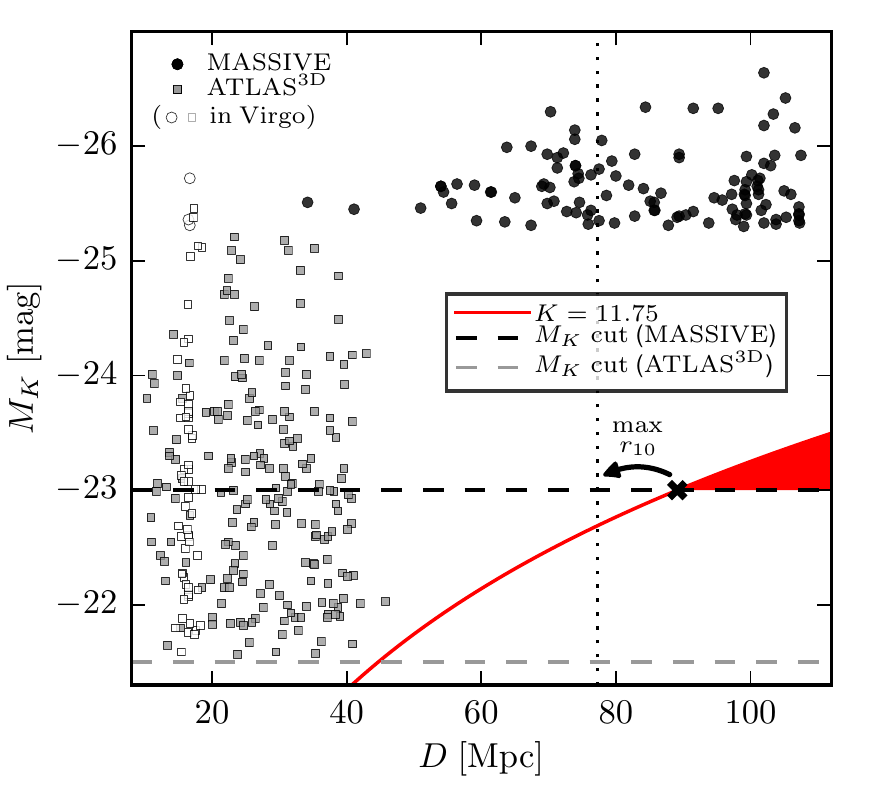}
\end{center}
\caption{Schematic of the magnitude limits of 2MRS (red line), the \atd\ sample (grey squares) and its parent sample (dashed grey line), and the MASSIVE sample (black circles) and its parent sample (dashed black line) defined for the purposes of calculating $\nu_{10}$.
  Galaxies in the MASSIVE volume that should be in the parent sample but are fainter than the 2MRS survey limit (red shaded region) may cause $\nu_{10}$ to be under-estimated.
  Because $r_{10}$, the distance to the 10$^{\rm th}$ neighbour, can be as large as $\sim$ 10 Mpc, this extends the potential impact of the incomplete region significantly beyond the intersection of the $M_K$ cut and $K=11.75$ mag to all galaxies to the right of the dotted line.
  Moving the cut for the MASSIVE parent sample up moves the dotted line to the right, meaning fewer galaxies impacted by incompleteness; however, it also causes more \atd\ galaxies to fall outside the cut.
  Our choice of $M_K < -23.0$ mag is a compromise between those competing effects.
}
\label{fig:nu-limit}
\end{figure}

In \citet{cappellarietal2011b}, the ten nearest neighbours were chosen from the \atd\ parent sample, containing all galaxies (not just ETGs) in the \atd\ volume with $M_K < -21.5$ mag.
This cut reflects the 2MRS survey limit of $K = 11.75$, which is illustrated in \mautoref{fig:nu-limit}.
Using the same $M_K$ cut to define a parent sample for MASSIVE would result in substantial incompletness; all galaxies between the red line and dashed grey line in \mautoref{fig:nu-limit} would be missing.
On the other hand, to guarantee zero impact from incompleteness to our $\nu_{10}$ calculation (moving the vertical dotted line all the way to the right of the figure) would require a cut of $M_K < -24.0$, which would cause most \atd\ galaxies to fall outside the cut entirely.
We want to make a fair comparison to \atd\ galaxies, so we recompute $\nu_{10}$ instead of using the values in \citet{cappellarietal2011b}, and thus want to keep more than a few galaxies inside our cut.
We choose to define the MASSIVE parent sample with $M_K < -23.0$ as a compromise between those two considerations.
This cut allows us to keep about half of the \atd\ galaxies for comparison, while about half of the MASSIVE galaxies have $\nu_{10}$ possibly impacted by incompleteness of the parent sample.

To estimate the impact of incompleteness on $\nu_{10}$, we repeated our calculation for a parent sample cut at $M_K < -24.0$.
Using only galaxies not impacted by incompleteness (i.e. those to the left of the dotted line in \mautoref{fig:nu-limit}), we found that expanding the parent sample from $M_K < -24.0$ to $M_K < -23.0$ results in a characteristic increase of $\Delta \log_{10} \nu_{10} \sim 0.6$.
This represents a worst case scenario for the bias in $\nu_{10}$ of galaxies to the right of the dotted line, since only those at the very edge of the volume experience the maximum effect of incompleteness.
Since $\nu_{10}$ covers six orders of magnitude, we judge this to be a minor impact.

Another difference between the MASSIVE and \atd\ surveys is the availability of accurate distance estimates (discussed in detail in \citealt{maetal2014} and \citealt{cappellarietal2011a} respectively).
For most of the MASSIVE galaxies, we use group distances from the catalogues of \citet{crooketal2007}.
This effectively flattens the galaxies in each group to the same distance, and would generally result in a higher $\nu_{10}$.
Although more accurate distances would result in more accurate values of $\nu_{10}$, we wish to make a fair comparison to \atd\ galaxies, so we do not use the accurate distances tabulated by the survey papers even when they are available.
Instead, we assign distances from the HDC catalogue first (if available), then from the LDC catalogue, and as a last resort use the raw 2MRS redshift distance.

\begin{figure}
\begin{center}
\includegraphics[page=2]{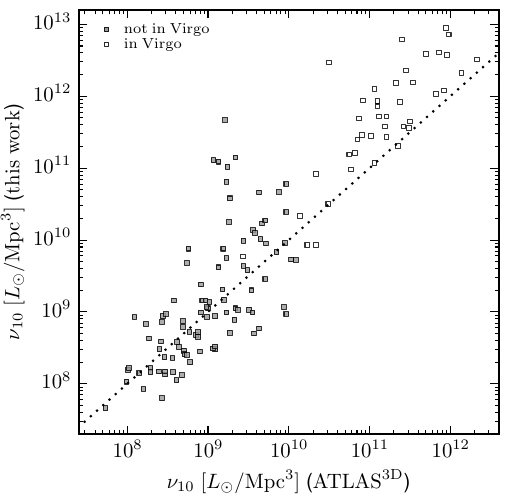}
\end{center}
\caption{
  Our recalculated $\nu_{10}$ versus the $\nu_{10}$ from \citet{cappellarietal2011b} for \atd\ galaxies.
  At low density, we see the effect of our more strict $M_K$ cut, resulting in a lower $\nu_{10}$ for most galaxies.
  Within groups and clusters, flattening each group to a single distance can result in somewhat reduced $r_{10}$, which results in significantly increased $\nu_{10}$ for a few galaxies.
}
\label{fig:nu-nu}
\end{figure}

\mautoref{fig:nu-nu} compares our recalculated $\nu_{10}$ to the values from \citet{cappellarietal2011b}.
Overall the agreement is reasonable, considering the two competing influences of our changes to the calculation.
First, we have a more strict $M_K$ cut on the parent sample, which will reduce $\nu_{10}$.
We see this at low densities, with reductions in $\nu_{10}$ up to an order of magnitude.
This is roughly in line with our comparison between the $M_K < -24.0$ and $M_K < -23.0$ cuts discussed above.
Second, we have ignored accurate distance estimates for nearby galaxies in favor of a more uniform assignment of group distances.
Flattening the groups to a single distance reshuffles the order of which neighbouring galaxies are closest, which may have a small impact on the total luminosity, but the major impact on $\nu_{10}$ comes from reduced $r_{10}$.
Even if the 10 neighbours are the same galaxies, $r_{10}$ is reduced to a 2-dimensional $R_{10}$ if all neighbours are in the same group.
It can be reduced further if galaxies that are nearly coincident on-sky, but are at opposite sides of the group along the line of sight, are counted as neighbours when they would not be otherwise.
A moderate change in $r_{10}$ has an impact of $r^3$ on the volume used to calculate $\nu_{10}$, and in a very few cases $\nu_{10}$ increases by up to 2 orders of magnitude.

The agreement between our new $\nu_{10}$ and the original values is good, considering the effects described above.
We also stress that enabling a fair comparison between MASSIVE and \atd\ galaxies is more important than increased accuracy of $\nu_{10}$ for individual galaxies.

\section{Bayesian error estimates for fraction of slow rotators}
\label{sec:err}

In \mautoref{sec:lam-pred} we compare the observed fraction of slow rotators $f_{\rm slow}$ as a function of environment to what we predict using $M_K$.
There is limited statistical power in certain bins, where the number of MASSIVE and/or \atd\ galaxies is small.
Thus we require a reasonable estimate of the error on $f_{\rm slow}$ so that we can make the comparison fairly and not overstate any differences.

For simplicity, we will ignore error bars on $\lambda_e$ and treat the classification of each individual galaxy as slow or fast as a 100\% certain measurement with no errors.
Although this is not true, the statistical errors due to sample size are our main concern.
The fraction of slow rotators must be between 0 and 1, and many simple estimates of the error are unsatisfactory.
(For example, a simple bootstrapping method would yield zero error for a subsample of 5 galaxies containing 0 slow rotators, even though there should be significant uncertainty due to the small sample size.)
Fortunately, our problem is equivalent to a well known example in Bayesian statistics, the problem of flipping a biased coin $N$ times and estimating the true probability of getting heads or tails.

For some fraction of coin flips (or slow rotators) $x$, the prior and posterior distributions can be conveniently defined by a Beta distribution:
\begin{equation}
  P(x) \propto x^{\alpha - 1} (1 - x)^{\beta - 1}
\end{equation}
with a mean of $\mu = \alpha/(\alpha + \beta)$.
The quantity $n = \alpha + \beta$ is often interpreted as the sample size, and the variance is $\mu(1-\mu)/(n+1)$.
The parameters of the posterior distribution, given a prior distribution and the measured numbers of fast and slow rotators, are $\alpha_{\rm post} = \alpha_{\rm prior} + N_{\rm slow}$, $\beta_{\rm post} = \beta_{\rm prior} + N_{\rm total} - N_{\rm slow}$.

We choose a prior distribution based on $f_{\rm slow}$ for each survey: $\mu_{\rm prior} = 0.78$ for MASSIVE and $\mu_{\rm prior} = 0.13$ for \atd, with $n_{\rm prior} = 5$ for both.
Then to obtain the error on $f_{\rm slow}$ in each specific bin of environment, we find the 68\% confidence interval of the posterior distribution.
These choices of prior are somewhat arbitrary (i.e. there is nothing special about $n_{\rm prior}=5$), but qualitatively give the behaviour we expect.
We have a weak prior assumption that any subsample of galaxies will have the same $f_{\rm slow}$ as the overall sample, so the errors will be slightly asymmetric towards that overall fraction, and the size of the error depends properly on the size of the subsample.
The slow fraction for \atd\ BGG galaxies in the bottom panel of \mautoref{fig:lam-bgg} is a good illustration of these properties.

\section{Comparing to \atd\ densities}
\label{sec:atd}

\begin{figure}
\begin{center}
\includegraphics[page=1]{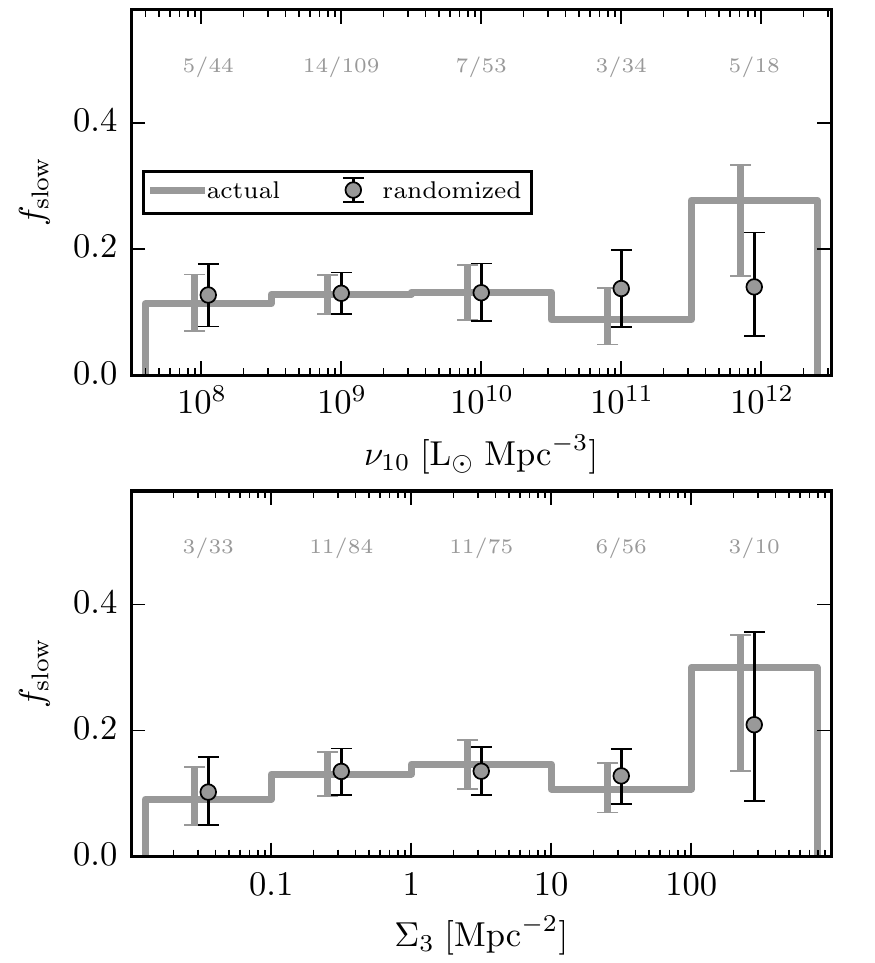}
\end{center}
\caption{Slow rotator fraction $f_{\rm slow}$ versus $\nu_{10}$ and $\Sigma_3$ for \atd\ galaxies, using densities originally tabulated in \citet{cappellarietal2011b}.
  While the observed $f_{\rm slow}$ appears similar in both cases, rising in the highest density bin, comparing to $f_{\rm slow}$ in the randomized sample (see \mautoref{sec:lam-pred}) shows an important difference.
  The slow fraction as a function of $\Sigma_3$ is nearly identical when randomized within bins of $M_K$, but for $\nu_{10}$ the randomized sample underpredicts $f_{\rm slow}$ in the highest bin.
}
\label{fig:slow-atd}
\end{figure}

\mautoref{fig:slow-atd} compares the observed slow rotatator fraction $f_{\rm slow}$ to the a test sample constructed by randomizing the slow/fast assignment of galaxies within bins of $M_K$.
(See \mautoref{sec:lam-pred} for details.)
This is similar to the results shown in \mautoref{fig:lam-pred}, but uses the local densities $\nu_{10}$ and $\Sigma_3$ tabulated in \citet{cappellarietal2011b} for the \atd\ sample.
The results in the top panel of \mautoref{fig:slow-atd} for the entire \atd\ sample, with $\nu_{10}$ calculated using the best available distance estimates, are slightly different from the results in the top right panel of \mautoref{fig:lam-pred}, calculated for galaxies with $M_K < -23.0$ using our simplified distances.

Both measures of local density find an increase in $f_{\rm slow}$ in the highest density bin, but the test samples (grey points in \mautoref{fig:slow-atd}) show an important difference.
The slow rotator fraction as a function of $\Sigma_3$ is well matched using the randomized test sample, but if the case of $\nu_{10}$ the test sample underpredicts $f_{\rm slow}$ at the highest bin.
This is very similar to the results for MASSIVE galaxies in the top right panel of \mautoref{fig:lam-pred}.


\bsp	
\label{lastpage}
\end{document}